\documentclass[a4paper,10pt]{article}
\usepackage{amsfonts}
\usepackage{amssymb}
\usepackage{amsmath}
\usepackage{amsthm}
\usepackage{latexsym}
\usepackage{layout}
\usepackage[english]{babel}
 
\usepackage{epsfig}
\usepackage{graphicx}
\usepackage{pdfsync}

\numberwithin{equation}{section} 
\newtheorem{theorem}{Theorem}[section]
\newtheorem{proposition}[theorem]{Proposition}
\newtheorem{lemma}[theorem]{Lemma}
\newtheorem{remark}[theorem]{Remark}
\newtheorem{definition}[theorem]{Definition}

\def\eeps{{E_{\eps}}}

\newcommand{\hide}[1]   {}

\newcommand{\dx}        {\,{\rm dx}}

\newcommand{\dist}        {\,{\rm dist}}
\newcommand{\eps}       {\varepsilon}
\newcommand{\dH}        {{\rm d}_{\mathcal H}}
\newcommand{\Huno}        {{\mathcal H}^1}

\newcommand{\Om}        {\Omega}
\newcommand{\R}         {\mathbb{R}}

\newcommand{\TTeps}  {{\mathcal T}_\eps}
\newcommand{\NNeps}  {{\mathcal N}_\eps}

\newcommand{\nada}[1]    {}
\newcommand{\ave}{-\hskip -.38cm\int}

\newcommand{\Per}{{\rm Per}}

\newcommand{\ii}{\texttt{i}}
\newcommand{\jj}{\texttt{j}}

\newcommand{\Z}{\mathbb Z}
\newcommand{\N}{\mathbb N}
\newcommand{\deps}{{\mathcal D}_\eps}

\newcommand{\sodue}{{{\mathcal SO}(2)}}
\def\admeps{{{\mathcal Ad}_\eps}(\Omega)}

\def\II{{\mathcal I}}
\def\dHuno{{\,d {\mathcal H}^1}}

\def\e{\eps}

\begin{document}

\title{Asymptotic analysis \\
of microscopic impenetrability constraints \\
for atomistic systems}

\author{
A. Braides and 
M.S. Gelli}
\date{}
\maketitle

\abstract\noindent In this paper we analyze a two-dimensional discrete model of nearest-neighbour Lennard-Jones interactions under the microscopical constraint that points on a lattice triangle maintain 
their order. This can be understood as a microscopical non-interpenetration constraint and amounts to 
the positiveness of the determinant of the gradient of the piecewise-affine interpolations of the discrete displacement. Under such a constraint we examine the continuum fracture energy deriving from  a discrete-to-continuum analysis at a scaling where surface energy is preponderant. We give a lower bound by an anisotropic Griffith energy. This bound is optimal if the macroscopic displacement satisfies some opening-crack conditions on the fracture site. We show that if such conditions are not satisfied then 
the computation of the energy due to continuum cracks may involve non-local effects necessary to bypass the positive-determinant constraint on crack surfaces and at points where more cracks meet. Even when the limit fracture energy may be described by a surface energy density, this may depend on the crack orientation both in the reference and in the deformed configuration. While these effects lead to very interesting analytical issues, they call into question the necessity of the determinant constraint for fracture problems.


\section*{Introduction}\label{secint}
In this paper we analyze on one hand continuum descriptions for 
fracture energies arising from discrete systems of particles linked by interatomic interactions,
and on the other hand computational problems deriving from microscopic constraints.
In the simplest model of atomistic interactions, the behavior of a collection of $N$ particles is governed by an energy that can be written as a sum of pair contributions; namely, it can be set in the form
\begin{equation}
E_N(\{u_i\})=\sum_{1\le i\neq j\le N} J(\|u_i-u_j\|),
\end{equation}
where $u_i$ is the position of the $i$-th atom, $\|u_i-u_j\|$ the distance between the corresponding pair of atoms, and $J$ is an interatomic potential, which is strongly repulsive at short distances and mildly attractive at long distances. The most common choice for $J$ is the {\em Lennard-Jones} potential 
\begin{equation}
J(r)={c_1\over r^{12}}-{c_2\over r^6}
\end{equation}
($c_1,c_2>0$). Note that in principle the total energy of the system is bounded from below by $N^2\min J$; however, a more refined estimate shows that it scales exactly as the number $N$ of atoms. This is in agreement with the intuition that ground states for $E_N$ arrange approximately in a regular lattice as $N$ increases ({\em crystallization}), so that the energy contribution of each particle is essentially described by the interaction only with its neighbours. At the same time it suggests that models of crystalline solids can be derived directly from such atomic interactions.


The crystallization for Lennard-Jones interactions in the general context describe above is still an open problem.
An important contribution has been recently given by Theil \cite{T}, who has studied a slightly weaker version of crystallization,
proving in two dimensions the optimality under compact perturbations of the ``minimal'' triangular lattice; i.e., the one for whose lattice spacing $\overline r$ it is minimal the average energy per particle
\begin{equation}\label{erg}
e(r)= \sum_{k\in\bf T\setminus \{0\}} J(r\|k\|),
\end{equation}
which describes the energy of a single particle in the lattice $r\bf T$ due to the interaction with the other particles, $\bf T$ being the unit triangular lattice.
Once crystallization is achieved, another important issue is whether it is maintained on states other than ground states; i.e., whether to (small) macroscopic deformations there corresponds uniform displacements at the atomic level ({\em Cauchy-Born rule}; see Friesecke and Theil \cite{FT}). 

In order to examine the behavior of atomistic systems far from ground states, under the hypothesis of crystallization we may consider the energy related to a density $\rho$. Noting that $\rho$ is proportional to $r^{-n}$, where $n$ is the space dimension (usually, $n=2,3$), this can be expressed as
$$
f(\rho)= \rho(C_1\rho^{12/n}- C_2\rho^{6/n}),
$$
where $C_1,C_2$ are determined by computing the energy of a single particle in a uniform lattice (in the case of a triangular lattice $f(\rho)=\rho\, e(r)$, where $e$ is defined in (\ref{erg}) and $r$ is the lattice spacing corresponding to the density $\rho$). 
This function $f$ is non-convex in an interval $(0,\rho_0)$, which highlights a {\em phase transition} at low densities, and suggests that large deformations involve a change in the crystalline structure which is instead achieved close to ground states.

From the standpoint of Continuum Mechanics, a description with an hyperelastic bulk energy is expected to hold close to ground states, while the same is expected to fail for a class of large deformations. In the one-dimensional case this can be achieved by introducing a fracture energy.
Following Truskinovsky \cite{trusk}, in this case, given $N$ particle positions $u_i$, these can be parameterized with $i=1,\ldots N$ in such a way that $u_i> u_{i-1}$, and write the energy as 
$$
E_N(\{u_i\})=\sum_{j>i} J(u_j-u_i).
$$
As $N$ increases ground states tend to arrange regularly on a lattice that we may suppose to be $\mathbb Z$; i.e., 
we may suppose that the one-dimensional energy per particle
$$
e(r)= \sum_{k\in{\mathbb N},\  k>0} J(rk),
$$
has its minimum at $\overline r=1$. In order to introduce a macroscopic deformation gradient, we can now scale and re-parameterize the same particles on $\e\mathbb Z$, where $\e={1\over N}$; i.e., $u_i= {1\over\e} u({i\e})$, 
so that they all can be seen as discretizations of functions defined on a single interval $[0,1]$. If we scale the energy as
$$
E_N(\{u_i\})=\sum_{j>i} J\Bigl({u(j\e)-u(i\e)\over\e}\Bigr).
$$
then the ground states are discretizations of the identity on $[0,1]$. 
%
If we let $N$ increase (or $\e\to0$), we may highlight two regimes:

({\em bulk scaling}) under the hypothesis of small perturbations $u(x)= x+\delta v(x)$ 
with small $\delta$,we have
$$
{u({(i+k)\e})-u({i\e})\over\e}\approx k+k\delta v'(i\e),
$$
so that, Taylor expanding $e$ at its minimizer $1$,
$$
E_N(\{u_i\})=N\min e+{N\over 2}e''(1)\delta^2\int_0^1|v'|^2\dx +o(\delta^2).
$$

({\em surface scaling}) if the macroscopic $u$ is discontinuous at a point parameterized by $i$ then we have an increase from the ground state energy of the order $J(+\infty)-\min e$, which gives a Griffith fracture energy at each such point.

This argument can be made rigorous, and it gives:

$\bullet$ elastic behavior close to ground states, with a linearized description given by the linearization of $J$ at ground states;

$\bullet$ a brittle fracture energy depending on the depth of the well of $J$ with respect to the infinity;

$\bullet$ opening fracture: the possibility of a parameterization with increasing $u$ implies that fracture may only open up, providing a natural non-interpenetration condition;

$\bullet$ surface relaxation: on external boundaries and on internal fracture sites the asymmetry of the atomic arrangements gives an additional surface term, highlighting a microscopic rearrangement close to those surfaces.

\smallskip
\noindent We note that all the features above can be included in a single analysis by choosing $\delta=N^{-1/2}$ in the notation above, so that (up to additive constants) bulk and surface terms have the same scaling. This has been done by Braides, Lew and Ortiz \cite{BLO}. We also remark that an analysis of local minima and of opening cracks suggests a cohesive fracture energy density of Barenblatt type, which is not directly given by the analyses above. However some cohesive theories can be shown to be ``equivalent'' to the ones obtained by the limit analysis (see Braides and Truskinovsky \cite{BT} and \cite{BLN13}).

The same study for the two or three-dimensional case presents greater difficulties, mainly because a natural parameterization with increasing functions is no longer possible. Hence, some simplifications have been made for this model with the scope of maintaining the features of the complete system of interactions and at the same time allow for an analytical study. A first assumption is to consider displacements as a perturbation from a ground state for which crystallization holds; i.e., in dimension two a perturbation from a state parameterized on the triangular lattice $\bf T$, or rather on a bounded portion $\Lambda$ of $\bf T$ (this corresponds to take $N$ as the numbers of points in $\Lambda$ in the previous notation). For such perturbation it makes sense to assume that only short-range interactions be taken into account. 
The range of such interaction can be indexed with a subset $S$ of ${\bf T}\setminus\{0\}$. The energies replacing $E_N$ can be written as
$$
F_\Lambda(u)=\sum_{k\in S} \sum_{i,j\in\Lambda, i-j=k} J_k(\|u_i-u_j\|),
$$
where $J_k$ represents the energy between points whose parameters $i,j$ differ by $k$ in the reference lattice. The simplest choice is considering only nearest-neighbour interactions on $\bf T$, with $S$ as the unit vectors of the triangular lattice, and $J_k$ independent of $k\in S$ a Lennard-Jones potential with minimum in $\overline r$. In this simplified model interactions are minimized when the corresponding $u_i-u_j$ are at distance $\overline r$, thus recovering uniform deformations. Unfortunately, such a simplified model admits many additional minimizers, as all deformations which are piecewise homotheties with ratio $\pm\overline  r$. In fact, if we compose a uniform deformation (e.g. a homothety) of ratio $\overline r$ with ``folding'' along a line of points in $\bf T$, the resulting nearest neighbours still are at the ``minimal'' distance $\overline r$. 

In order to prevent undesired ``foldings'' at a discrete level without considering longer-range interactions, Friesecke and Theil \cite{FT} proposed to add a three-point condition on neighbouring nodes. In the case of a triangular lattice, this condition simply amounts to requiring that
$$
\hbox{ det } \nabla u>0,
$$
where $u$ is the affine interpolation of the function defined on the vertices of each triangle. In this way only discretizations of homotheties with positive ratio equal to $\overline r$ are minimizers of the energy. In their paper, Friesecke and Theil treat the seemingly ``un-natural'' case of energies parameterized on a square lattice. Actually, when finite-range lattice energies are considered, the choice of the underlying lattice is a matter of convenience, and the square lattice is the simplest where to consider at the same time nearest neighbours and next-to-nearest neighbours to highlight the possibility of non-uniform ground states ({\em failure of the Cauchy-Born rule}). 

In this paper we treat a two-dimensional system of nearest-neighbour Len\-nard-Jones interactions with the positive-determinant constraint focusing on the surface scaling. We will use the terminology and techniques of $\Gamma$-convergence applied to a discrete-to-continuum analysis \cite{GCB}. In this framework we examine the overall behavior of the energies $F_\Lambda$ as the size of $\Lambda$ diverges, by considering $\Lambda={1\over\e} (\Omega\cap {\bf T})$, with $\Omega$ a fixed bounded domain in $\mathbb R^2$, and using $\Omega\cap\e \bf T$ as the set of parameters.
The scaled energies we are going to examine will be of the form
$$
F_\e(u)=\sum_{i,j\in\Omega\cap\e {\bf T}, |i-j|=\e}\e \,J\Bigl(\Bigl\|{u_i-u_j\over\e}\Bigr\|\Bigr),
$$
where $u_i$ is the value of the discrete function $u$ at the node $\e i\in \Omega\cap\e {\bf T}$, and the piecewise-affine interpolation of $u$ on the triangulation related to $\e\bf T$ is supposed to satisfy the positive-determinant constraint.
The scaling $\e$ heuristically can be explained by considering, as in the one-dimensional case, the contribution of a set of indices $I$ where $\|u_i-u_j\|>\!>1$ and noting that under this scaling the finiteness of the energy asymptotically implies that they have the dimension of a line, so that they can be regarded as an interface. Under these assumption we will address the two issues

$\bullet$ determine whether some condition of ``opening crack'' still hold in the two-dimensional case;

$\bullet$ characterize a limit continuum surface energy defined on functions defined on $\Omega$.

The other two issues present in the one-dimensional analysis; i.e., the characterization of the bulk energy close to ground states and surface relaxation have been separately addressed in slightly different hypotheses by Braides, Solci and Vitali \cite{BSV} (for the bulk analysis) and Theil \cite{Th2} (for the external surface relaxation). An analysis of the small-displacement regime has been carried over by Friedrich and Schmidt \cite{F1,FS1,FS2}.
\begin{figure}[htbp]
\begin{center}
\includegraphics[width=.45\textwidth]{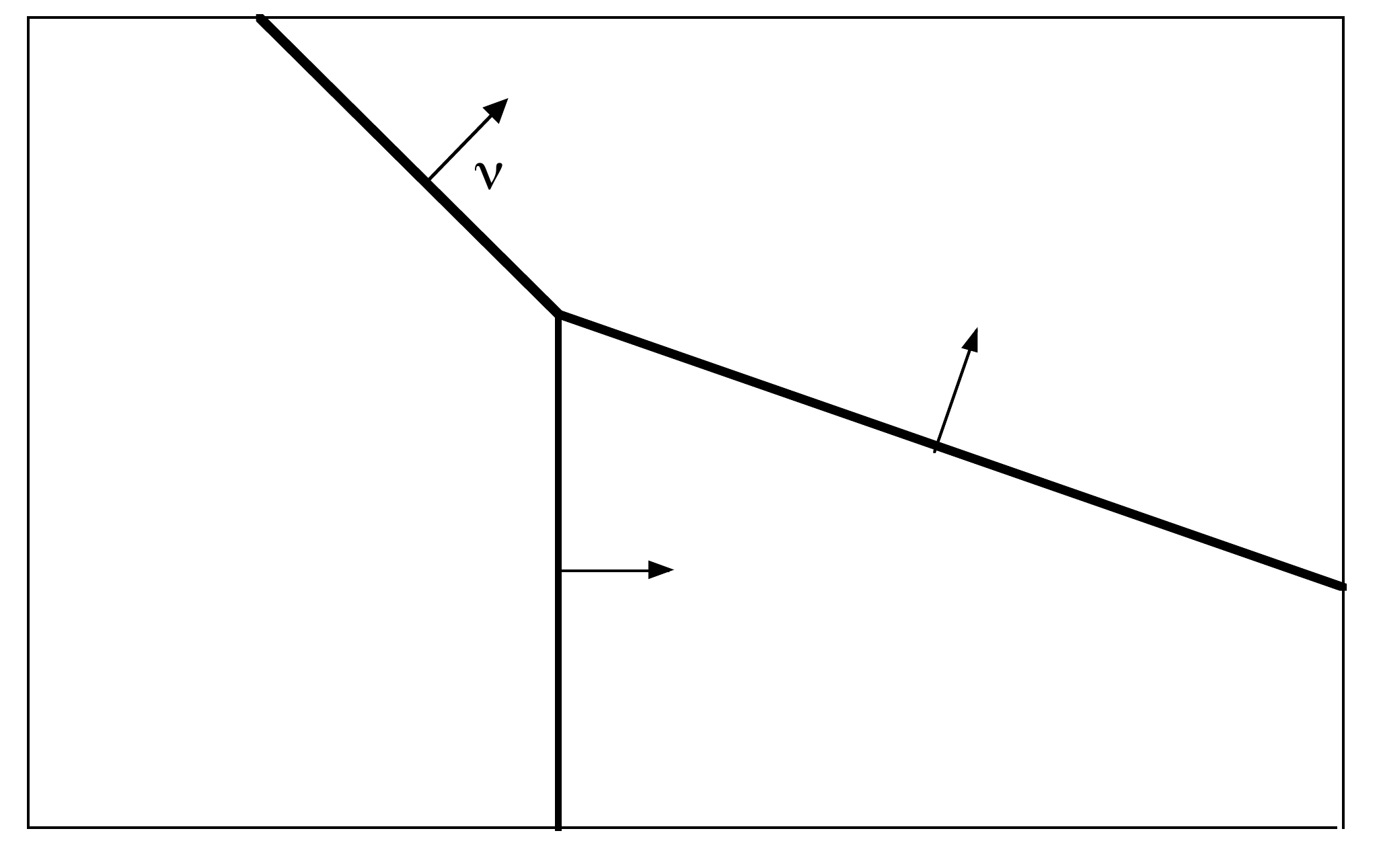}
\includegraphics[width=.45\textwidth]{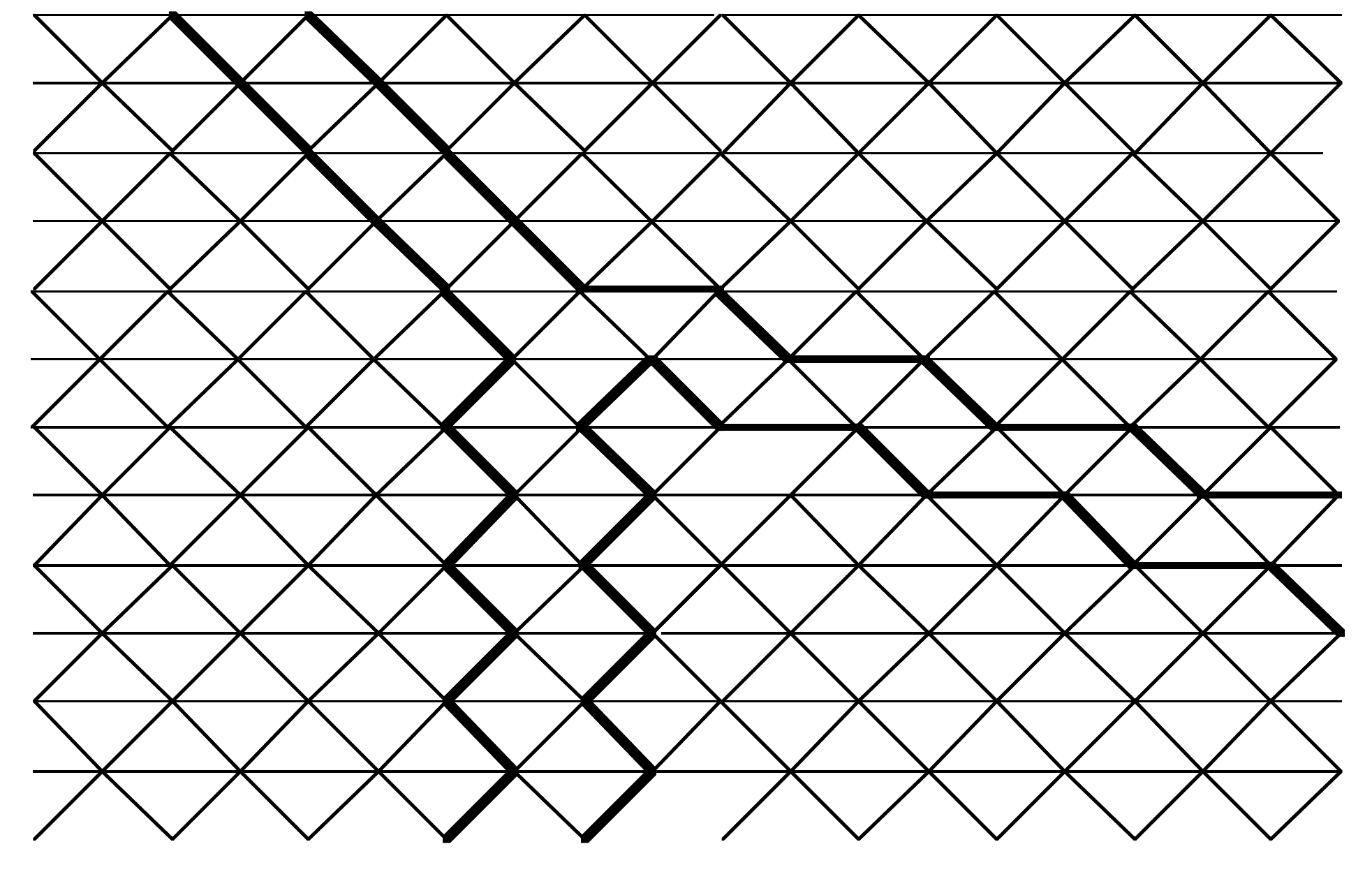}
\caption{A reference configuration with fracture site and its macroscopic normals, and its underlying triangulation}
\label{illustration1-2}
\end{center}
\end{figure}

In order to makes the analysis clearer we will scale the energies by an additive constant as
$$
F_\e(u)=\sum_{i,j\in\Omega\cap\e {\bf T}, |i-j|=\e}\e\Bigl( \,J\Bigl(\Bigl\|{u_i-u_j\over\e}\Bigr\|\Bigr)-\min J\Bigr),
$$
so that the energy density is always positive. As a first remark we will note that, using a result by Chambolle, Giacomini and Ponsiglione \cite{CGP}, gradients of limits of sequences $(u_\e)$ with equi-bounded energy are piecewise rotations with an underlying partition of $\Omega$ into sets of finite perimeter. This allows us to focus on the boundaries of such sets, where we have a normal $\nu$ on whose two sides we have the values $u^\pm(x)$ of $u$ and two rotations $R^\pm$ among those labeling the sets of the partition. Note that from the standpoint of the microscopical triangulations the interfaces are the limits of triangles of side-length $\e$ which are deformed by $u_\e$ into triangles with one side (actually two) of diverging length but with the same ordering of the vertices (this corresponds to the positive-determinant constraint). 
\begin{figure}[htbp]
\begin{center}
\includegraphics[width=.45\textwidth]{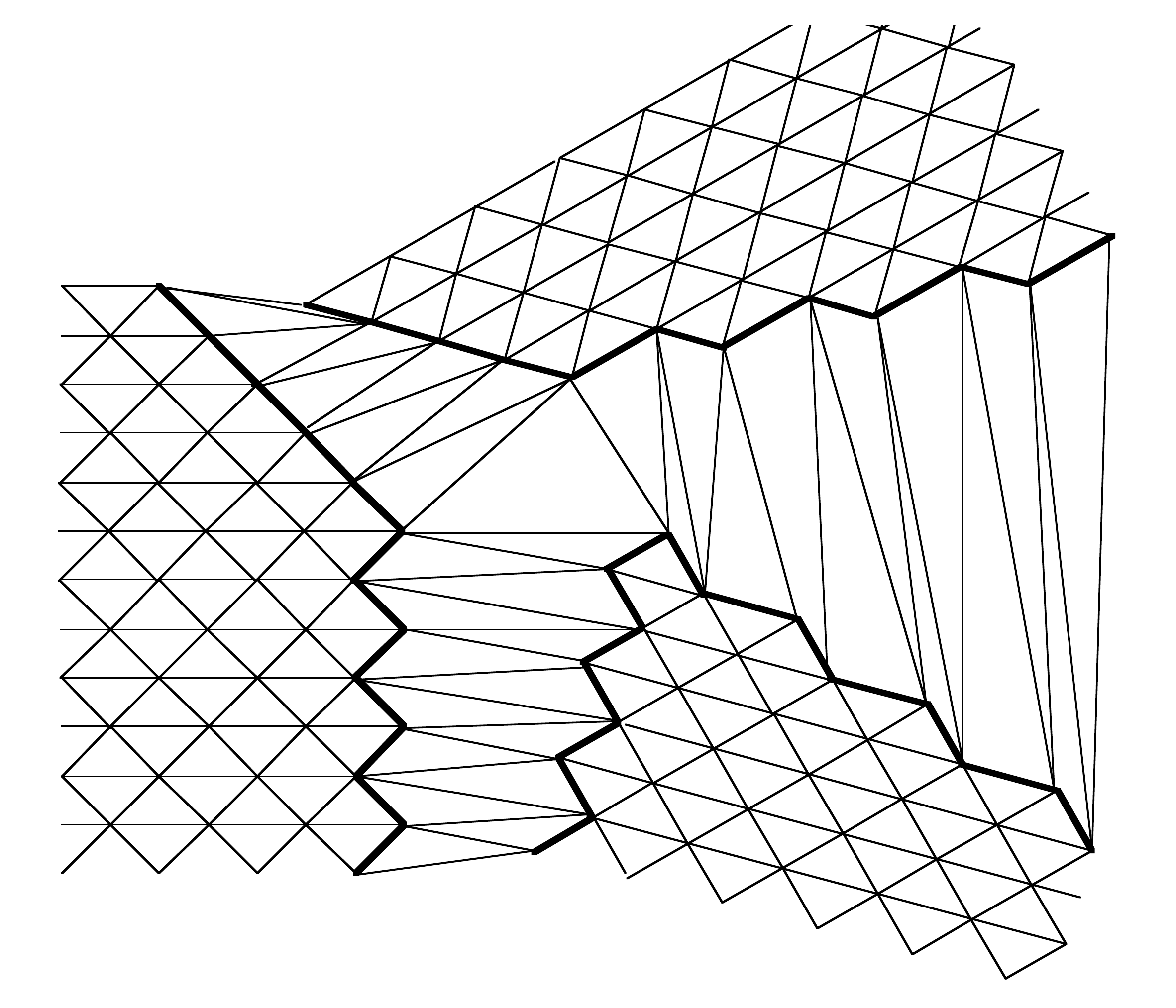}
\includegraphics[width=.5\textwidth]{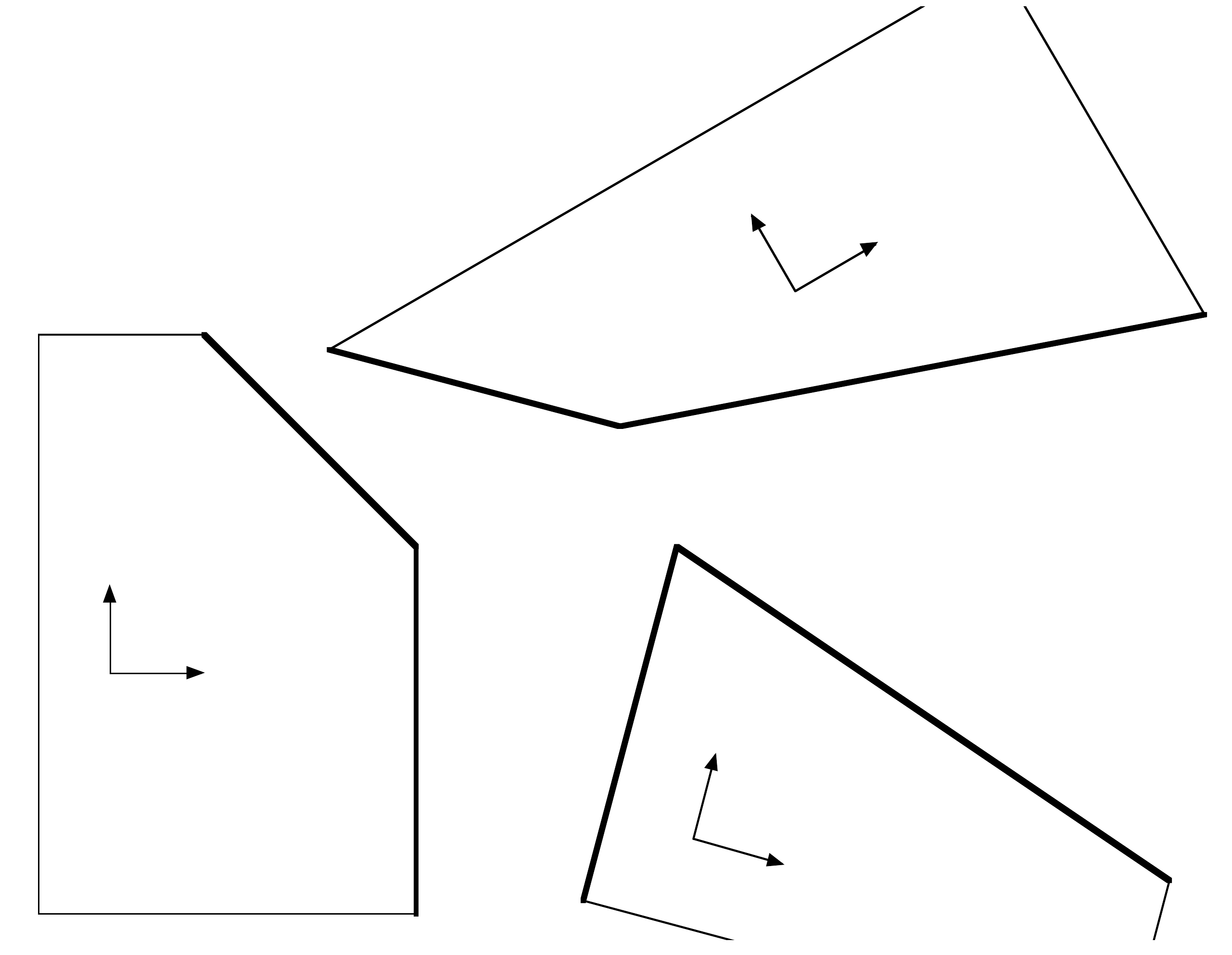}
\caption{A deformed microscopical configuration satisfying the determinant constraint, and its macroscopic limit with the corresponding rotations}
\label{illustration3-4}
\end{center}
\end{figure}
If only a layer of triangles is deformed that gives a limit interface, then this gives a relation between $\nu$, $u^\pm(x)$ and $R^\pm$. In the simplest case when $\nu$ is orthogonal to one of the unit vectors of $\bf T$ (we call those vectors {\em coordinate normals}), this relation reads as
$$
\langle u^+(x)- u^-(x), R^\pm \nu \rangle\ge 0\,.
$$

This can be regarded as an {\em opening-fracture condition} in the ``finite'' case: note that in the ``infinitesimal case'' when $u$ is a small variation of a ground state; i.e., $u=id +\delta v$ with $\delta<\!<1$, then this condition reduces to the usual {\em infinitesimal crack-opening condition}
$$
\langle v^+(x)- v^-(x), \nu \rangle\ge 0\,
$$
since $R^\pm = id +o(1)$. If $\nu$ is not a coordinate normal then  the opening-crack condition gets more complicated, due to microscopic anisotropies of the lattice, which disfavor cracks not orthogonal to lattice directions. The same anisotropies appear, as usual for lattice systems, in the form of the surface energy density.
However, the situation described above is not the only possibility, since more than one layer of triangles may be ``strongly deformed''. This gives a higher energy on the interface, but relaxes the constraints on $\nu$, $u^\pm(x)$ and $R^\pm$. Moreover, additional energy contributions may be given by points where three or more interfaces meet; in this case, even though the opening-fracture condition above may be satisfied on each interface, the system of interfaces may be incompatible with the positive-determinant condition in their common point.  It must be observed that by introducing an high number of extra interfaces at the discrete level, the finite opening-fracture condition can be removed altogether, at the expenses of a complex non-local form of the final energy. 

We finally note that, even when non-local effects are neglected, the fracture energy density depends 
not only on the crack orientation $\nu$ and the crack opening $u^+-u^-$, but also on the image of the
crack in the deformed configuration, described by the tangential derivatives of $u^\pm$ on both sides of the crack. Such types of energies seem to be of interest in themselves, and the corresponding analytical techniques still relatively undeveloped (see \cite{ABG}).

\section{Some notation and preliminary results} 

In the sequel we denote by $\langle\cdot,\cdot\rangle$ the scalar product in $\R^2$ and 
with $|\cdot|$ the usual euclidean norm, both for vectors in $\R^2$ and matrices in 
$\mathcal M (2\times 2;\R)$, the meaning being clarified by the context. Moreover we will also use 
the symbol $\langle A, y\rangle$ to denote the duality pair for $A\in \mathcal M(2\times2;\R)$ 
and $y\in\R^2$. For every $x,y\in \R^2$, $[x,y]$ denotes the segment joining $x$ and $y$. 

The functional space involved in our discrete-to-continuous analysis will consist of measurable 
vector-valued functions whose components are {\itshape special functions 
with bounded variation}. Such functions are commonly used in the variational theories of Fracture
(see e.g. the seminal paper by Francfort and Marigo \cite{FM},
the book by Bourdin, Francfort and Marigo \cite{BFM} and the review paper \cite{AB}).

\begin{definition} Given $\Omega\subset \R^2$ a fixed bounded open set,
a function $u\in L^1(\Om,\R^2)$ is a {\rm function with bounded variation} in $\Om$, denoted by $u\in BV(\Om, \R^2)$, if its distributional derivative $Du$ is a vector-valued Radon measure on $\Omega$. 

We say that $u$ is a {\rm special function with bounded variation} 
in $\Om$, and we write $u\in SBV(\Om, \R^2)$, if $u\in BV(\Om, \R^2)$ and its distributional derivative $Du$ can be represented on any Borel set $A\subset \Omega$ as 
$$
Du(A)=\int_A\nabla u (x) \, dx +\int_{A\cap S_u} (u^+(x)-u^-(x))\otimes \nu_u(x) \, d\Huno (x).
$$ 
for a countably $\Huno$-rectifiable set $S_u$ in $\Omega$ that coincides $\Huno$-almost everywhere with the complement in $\Omega$ of the Lebesgue points of $u$. Moreover, $\nabla u (x)$ is the approximate gradient of $u$ at $x$, $\nu_u(x)$ is a unit normal to $S_u$, defined for $\Huno$-almost every  $x$, and $u^+(x),u^-(x)$ are the traces of $u$ on both sides of $S_u$ (according to the choice of $\nu_u (x)$). Here the symbol $\otimes$ stands for the tensorial product of vectors; i.e., for any $a,b\in \R^2$ $(a\otimes b)_{ij}:=a_ib_j$.  
\end{definition}

For fine properties of $BV$ and $SBV$ functions and a rigorous definition of all the quantities introduced above we refer to \cite{AFP}, Ch.~4. 
SBV functions enjoy a good compactness property 
that here is stated in its simplest form, suited for our functionals. 

\begin{theorem}\label{compactnessgsbv}
Let $\left(u_n\right)\subset SBV(\Om, \R^2)$ be satisfying 
\begin{equation}\label{clms}
\sup_n\left\{ \int_\Om \left|\nabla u_n\right|^p\dx+\|u_n\|_{L^\infty (\Omega, \R^2)}+
\Huno(S_{u_n})\right\}<+\infty, 
\end{equation}
for some $p>1$. 
Then there exist a subsequence $ u_{n(k)}$ and a function $u\in SBV(\Om, \R^2)$ such that $u_{n(k)}\to u$ in measure. 
In addition $D^a u_{n(k)}\to D^a u$ and $D^j u_{n(k)}\to D^j u$ weakly* as measures. 
\end{theorem} 

As a consequence we easily get that the limit function $u$ in the previous theorem 
has the further property that $\int_\Om \left|\nabla u_n\right|^p\dx+\|u\|_{L^\infty (\Omega, \R^2)}+\Huno(S_u)<+\infty$. 

Moreover estimate \eqref{clms} allows to discuss separately the lower semicontinuity of suitable bulk and surface integral 
involving, respectively, the absolutely continuous and jump part of the generalized distributional derivative.  

In order to treat lower semicontinuous bulk integrals in the vectorial setting it is necessary to recall the notion of 
quasiconvexity and quasiconvex envelopes. 

\begin{definition}\label{quasiconvex}
We say that a lower-semicontinuous integrand $\psi:\R^{2\times 2}\to 
\R\cup\{\pm\infty\}$ is \emph{quasiconvex} if
\begin{equation}\notag
\psi\left(A\right) {\mathcal L}^2\left( \Om \right) 
\leq\int_\Om \psi\left(A+\nabla\varphi(x)\right)\dx  
\end{equation}
for every $A\in\R^{2\times 2}$ and $\varphi \in W^{1,\infty}_0\left(\Om,\R^2\right)$. 
In addition, for any $f:\R^{2\times 2}\to \R\cup\{\pm\infty\}$, the 
\emph{quasiconvex envelope $f^{qc}$} of $f$  is defined as 
$$
f^{qc}(A):=\sup\left\{g(A)\,:\, g\  quasiconvex, \  g\le f\right\}.
$$  
\end{definition}

In case $f$ is real valued and continuous one can give a precise description of its quasiconvex envelope.

\begin{proposition}\label{envelopeform}
Let $f:\R^{2\times 2}\to \R$ be continuous. Then 
\begin{equation}\label{qcenvelope}
f^{qc}(A)=\inf\left\{\ave_\Om f\left(A+\nabla\varphi(x)\right)\dx\,:\, \varphi \in W^{1,\infty}_0\left(\Om,\R^2\right) \right\}	
\end{equation}
for $A\in\mathcal M(2\times 2;\R)$.
\end{proposition}

The following theorem deals with lower semicontinuity along sequences of  SBV functions weakly converging \cite{AFP}.   

\begin{theorem}\label{semicontinuity}
Let $\left(u_n\right)\subset SBV(\Om, \R^2)$ be satisfying 
\begin{equation}\label{clqcvx}
\sup_n\left\{ \int_\Om \left|\nabla u_n\right|^p\dx+
\Huno (S_{u_n})\right\}<+\infty 
\end{equation}
for $p>1$.
If $u_n\to u$ in measure, then $u\in SBV(\Om, \R^2)$. 
Moreover, for every quasiconvex integrand 
$\psi:\R^{2\times 2}\to \left[0,+\infty\right)$ such that 
$$
\left|\psi\left(X\right)\right|\leq C\left(1+\left|X\right|^p\right)
$$
for every $X\in\R^{2\times 2}$ and $C$ a positive constant, and for any norm $\varphi$ in $\R^2$, 

there holds
\begin{equation}\label{lscbulk}
\int_\Om \psi\left(\nabla u\right)\dx\leq 
\liminf_n\int_\Om\psi\left(\nabla u_n\right)\dx,  
\end{equation}

\begin{equation}\notag
\int_{S_u}\varphi(\nu_u)\dHuno\leq\liminf_n	\int_{S_{u_n}}\varphi(\nu_{u_{n}})\dHuno. 
\end{equation}
\end{theorem}

\bigskip
The final tool we present is a `slicing result' that allows to treat $2$-dimensional energies  
by reducing to $1$-dimensional ones \cite{AFP}.
  
Before entering into details we introduce some notation. For $\xi\in
S^1$ let $\Pi^\xi:=\{y\in\R^2 :  \langle y, \xi\rangle =0\}$ be the line through the origin  
orthogonal to $\xi$.  If $y\in \Pi^\xi$ and $E\subset \R^2$ we set
\begin{equation}\label{sezione1} 
E^{\xi,y}:=\{t\in\R:y+t\xi\in E\}.
\end{equation}
Moreover, if $u:E\to\R^2$ we define the function $u^{\xi, y}:E^{\xi,y}\to\R^2$
by
\begin{equation}\label{sezione2}
u^{\xi, y}(t):=u(y+t\xi).
\end{equation}

\begin{theorem}\label{slicing}
{\rm (a)} Let $u\in SBV(\Omega, \R^2)$. Then, for all $\xi\in S^1$ the
function $u^{\xi, y}$ belongs to $SBV(\Omega^{\xi,y},\R^2)$ for
$\Huno$-almost every $y\in\Pi^\xi$. Moreover for such $y$ we have
$$
u^{\xi, y}(t)=\langle\nabla u(y+t\xi ),\xi\rangle\quad\hbox{ for almost every } t\in
\Omega^{\xi,y},
$$
$$
S(u^{\xi, y})=\{t\in\R:y+t\xi\in S_u\},
$$
$$
u^{\xi, y}(t\pm)=u^{\pm}(y+t\xi)\quad or\quad u^{\xi, y}(t\pm)=u^\mp
(y+t\xi),
$$
according to the cases $\langle\nu_u,\xi\rangle>0$ or
$\langle\nu_u,\xi\rangle<0$ (the case $\langle\nu_u, \xi\rangle=0$ being
negligible) and for all Borel functions $g$
$$
\int_{\Pi^\xi}\sum_{t\in S(u^{\xi, y})}g(t)\, d\Huno
(y)=\int_{S_u}g(x)|\langle\nu_u,\xi\rangle |\,d\Huno.
$$
\indent
{\rm (b)} Conversely, if $u\in L^\infty(\Omega,\R^2)$ and, for all $\xi\in
\{e_1,e_2\}$ and for $\Huno$-almost every~$y\in\Pi^\xi$, $u^{\xi, y}\in
SBV(\Omega^{\xi,y}, \R^2)$ with 
$$
\int_{\Pi^\xi}\Bigl(\int_{\Omega^{\xi,y}}|\dot u^{\xi, y}|^p+\#(S(u^{\xi,
y}))\Bigr)\,
d\Huno(y)<+\infty,
$$
then $u\in SBV(\Omega, \R^2)$.
\end{theorem}

\bigskip

We introduce now a generalization of a rigidity type result in the SBV setting due to Chambolle, Giacomini,
 Ponsiglione (see \cite{CGP}). Before going through the statement we need some preliminary definitions. 
 First we recall the notion of Caccioppoli partition of a domain (see e.g.~\cite{CT}).  

\begin{definition}
We say that a collection of pairwise disjoint measurable sets $\{E_h\}_{h\in\N}$ is a \emph{Caccioppoli partition of $\Omega$}  
if $\cup_{h\in\N}E_h=\Omega$ and 
$$
\sum_{h\in\N} \Per (E_h)<+\infty.
$$ 
Moreover, given any rectifiable set 
 $K\subset\Omega$ we say that a \emph{Caccioppoli partition of $\Omega$ is subordinated to $K$} 
 if $\Huno(K\setminus\cup\displaystyle_{h\in\N} \partial^* E_h)=0$.  
 \end{definition}

The following result is the $2$-dimensional version of the one contained in \cite{CGP}. 
\begin{theorem}\label{rigidity} 
Let $u\in SBV(\Omega,\R^2)$ with $\Huno(S_u)<+\infty$ and $\nabla u(x)\in \sodue$ for almost every $x\in\Omega$. 
Then there exists a Caccioppoli partition $\{E_h\}_{h\in\N}$ of $\Omega$ subordinated to $S_u$ such that 
for almost every $x\in\Omega$ 
$$u(x)=R_h x+q_h\hbox{ on } {E_h}$$
where $R_h\in\sodue$ and $q_h\in\R^2$.
\end{theorem}

\section{Formulation of the problem}\label{secfor}

In the sequel $\Omega$ will be any open bounded open set in $R^2$.  
With fixed discretization step $\eps >0$ the reference lattice is given by 
$L_\eps:=\eps\langle \eta^1, \eta^2\rangle_{\Z^2} $ 
where $\eta^1=(1,0),\eta^2=(1/2,\sqrt{3}/2)$. We introduce also the following notations
$$
\eta^3=\eta^1-\eta^2, \qquad \textsc{S}=\{\pm\eta^1,\pm\eta^2,\pm\eta^3\}.
$$
Note that $\textsc{S}$ is the set of unitary vectors in the lattice $L_1$ and for each 
$\ii\in L^1$ $\ii +\textsc{S}$ is the set of its nearest neighbours. 
 
We define also the set $\textsc{D}$ of {\em coordinate directions}  as 
$$
\textsc{D}=\{\eta^\perp\,:\, \eta\in \textsc{S} \}.
$$

Dropping the dependence on $\eps$ whenever no confusion may arise 
we will use the symbol $\textsc{T}$ to denote any triangle with vertices in $L_\eps$ 
and sides of length $\eps$ and ${\mathcal T}_\eps$ will denote the sets of all such triangles. 
As already pointed out in the 
Introduction the choice of the lattice relies on the fact that $L_1$ is the 
simplest Bravais lattice in dimension $2$ 
compatible with a Cauchy-Born hypothesis. 
Before introducing the precise definition of the functionals object of our analysis  
we list here some notation used in the following sections for sets of indices or triangles: 
$$
\NNeps (\Omega)=\{(\ii,\jj)\in L_\eps\times L_\eps\,:\, [\ii,\jj]\subset \Omega, \, 
|\ii-\jj|=\eps, \ii\prec\jj \}, 
$$  
$$\TTeps^c (\Omega)=\{\textsc{T}\in \TTeps\,:\, \textsc{T}\subset \Omega\},
$$
where the symbol $\ii\prec\jj $ stands for the standard lexicographic order in $\R^2$. 
Thanks to this choice any pair of nearest-neighbouring indices is counted only once 
in the energy contribution. 

With this notation given a discrete vector-valued displacement $u:L_\eps\cap\Omega\to \R^2$
we consider its associated energy     
$$
\sum_{ (\ii,\jj)\in \NNeps (\Omega)}\eps J\Big(\Big|\dfrac{u(\ii)-u(\jj)}{\eps}\Big|\Big) 
$$
where $J:[0,+\infty)\to [0,+\infty)$ is a continuous function satisfying the following structure properties:
\begin{itemize}
	\item[(i)] $\min J=J(z_0)=0$;
	\item[(ii)] $\displaystyle\lim_{z\to+\infty}J(z)=J_\infty>0$;
	\item[(iii)] $\displaystyle\lim_{z\to 0+}J(z)=+\infty$;
	\item[(iv)] for any $\delta>0\, \exists\, c_\delta$ such that $J(z)\ge c_\delta (z-z_0)^2$ for $|z|\le\delta$. 
\end{itemize} 
 
In what follows for the sake of computational simplicity we simply assume $z_0=1$. 
Clearly, up to a scaling argument, the analysis remains valid for the general case. 

\smallskip
As customary, in order to pass from discrete systems to a continuum formulation, it is convenient to identify 
a function $\{u(\ii)\}_{\ii\in L_\eps\cap\Omega}\subset \R^2$ with an element of $L^1(\Omega,\R^2)$.

\begin{definition}[interpolation]
Given a discrete vector-valued function $u:L_\eps\cap\Omega\to \R^2$ we 
define its {\em interpolation} in the whole $\Omega$ as a function coinciding 
on each triangle $\textsc{T}\in\TTeps^c(\Omega)$ with the linear interpolation of the 
values in its vertices.    
\end{definition}

Note that an interpolation is not uniquely defined on triangles close to $\partial\Omega$.
This will not affect our arguments since convergences are always considered in the interior of $\Omega$.
We will identify a discrete function with its interpolation and maintain the same 
notation for both the discrete and continuous version, the notation  being
clarified by the context. 

\begin{remark} \rm
In this setting another common procedure is to identify a discrete function with a piecewise-constant element in $L^1(\Omega, \R^2)$ by  assigning constant value $u(\ii)$ on the cell $\{\ii+\eps (\lambda\eta^1+\mu \eta^2)\, :\, \lambda, \mu \in [0,1]\}$, 
$\ii\in L_\eps \cap\Omega $. In fact, none of the results stated in the following would be affected by this choice (see the discussion contained in \cite{AFG}). Considering discrete functions as continuous piecewise-affine ones allows us to formulate the orientation preserving constraint 
in terms of the standard determinant of $\nabla u$.  
\end{remark} 
We are now able to introduce the class of admissible functions; i.e., a proper  
class of vector-valued functions on the lattice $L_\varepsilon$ that are 
\emph{orientation preserving}: 
$$
\admeps =\{u :L_\eps\cap \Omega\to\R^2\,:\,\det (\nabla u)> 0 \hbox{ a.e. in } \TTeps^c (\Omega)\}.
$$ 
Finally, we define the functional $\eeps$ on $L^1(\Omega, \R^2)$ as 
\begin{equation}\label{energies}
\eeps(u)=\begin{cases}\displaystyle\sum_{(\ii,\jj)\in \NNeps (\Omega) }\eps J\Big(\Big|\dfrac{u(\ii)-u(\jj)}{\eps}\Big|\Big) & \hbox{ if }\,u\in\admeps \\
+\infty  & \hbox{ if } u\in L^1(\Omega,\R^2)\setminus\admeps .\\
\end{cases}	
\end{equation}

\bigskip

\begin{remark} \rm
Note that energies defined in \eqref{energies} account only for interactions well contained in $\Omega$ and the orientation-preserving constraint is not imposed a priori on an affine extension of u on the triangles intersecting $\partial \Omega$. 
Actually, if one is interested in minimum problems 
endowed with boundary conditions (together with some perturbation or fidelity terms), a standard procedure  
in the framework of $\Gamma$-convergence is to focus on the `principal' part of the total energy, neglecting at first boundary data. 
Once this task is accomplished one can further deal with the initial problems up to modifying the limit energy 
according to the contribution arising from recovery sequences with correct boundary datum.    
\end{remark}

To proceed further with our analysis we need to fix a convergence on $L^1(\Omega, \R^2)$. 

\begin{definition}[discrete-to-continuum convergence]\label{convergenza} According to the our identification of discrete functions with \emph{interpolations}, given $u_\eps, u\in L^1(\Omega, \R^2)$ we say that $u_\e\to u$, and we write simply $u_\eps\to u$, if we have  $\sup_\eps \|u_\eps\|_{L^\infty (\Omega, \R^2)} <+\infty $ and $\lim_{\e\to 0^+} \|u_\eps-u\|_{L^1 (\Omega, \R^2)}=0$.
\end{definition}

\section{Compactness and a lower bound}\label{lobaloba}
In this section we first provide a description of the domain of any $\Gamma$-limit functional of the discrete energies defined in \eqref{energies}.  
As already mentioned in the Introduction this space consists of piecewise rigid deformations $u$ with finite crack energy in the sense of Griffith's theory; i.e., $\Huno (S_u)<+\infty$.  

As a second result we exploit  geometric measure theory techniques to establish a lower-bound estimate for the $\Gamma$-$\liminf_\e\eeps (u)$ without imposing any a-priori hypothesis on the deformation $u$. 
Note that in the next section this bound will be proved to be optimal for the class of piecewise rigid deformations `with opening fracture' in the sense of the anisotropies of the reference lattice. For such deformations the limit fracture energy is simply governed by an anisotropic Griffith-type energy density
which reflects the anisotropies of the underlying triangular lattice.

\begin{proposition}\label{compactness}
Let $\{u_\eps\}\subset\admeps$ be such that $\liminf \eeps (u_\eps)<+\infty$ and 
\begin{equation}\label{controlloL1}
\sup_\eps \|u_\eps\|_{L^\infty (\Omega, \R^2)} <+\infty . 
\end{equation} 
Then there exists a Borel function $u$ such that, up to subsequences, $u_\eps\to u$ in $L^1 (\Omega, \R^2)$. Moreover, 
\begin{itemize}
	\item[{\rm(i)}] \emph{(finite Griffith fracture energy)} $u\in SBV (\Omega, \R^2)$ with $\Huno (S_u)<+\infty$; 
	
	\item[{\rm(ii)}] \emph{(piecewise rigidity)} there exists a Caccioppoli partition $\{E_h\}_{h\in\N}$ subordinated to $S_u$ such that for almost every $x\in E_h$ we have  $u(x)=R_h x+q_h$ for suitable $R_h\in\sodue$ and  $q_h\in\R^2$. 
	\end{itemize}
\end{proposition}
\proof 
As a first step we observe that any pair $(\ii,\jj)\in \NNeps (\Omega)$ belongs to two different triangles having 
the segment $[\ii,\jj]$ as a side. Hence if we take into account a factor $1/2$ we may estimate the energies $\eeps (u)$ from below with the integral functionals obtained as a superposition of gradient energies indexed  by the triangles $T$ varying in $\TTeps^c (\Omega)$. 
Actually, for any $B$ open set compactly supported in $\Omega$ and for any $v\in\admeps$, we have 
\begin{eqnarray} \label{stima1}\nonumber
\eeps (v)&\ge& \dfrac{\eps}{2|T|}\sum_{T\in\TTeps^c (\Omega)}\sum_{k=1}^3\int_T  J (|\langle \nabla v, \eta^k\rangle|)\dx\\
&\ge& \dfrac{2}{\sqrt{3}\eps}\int_B\sum_{k=1}^3 J (|\langle \nabla v, \eta^k\rangle|)\dx. 
\end{eqnarray}
According to this standpoint, we are led to considering integral functionals with energy density 
$\hat J(A):=\sum_{k=1}^3 J (|\langle A, \eta^k\rangle|)$.  
Hence, with fixed any $s\in(0,1)$ we distinguish `good' or `bad' triangles $T$ depending whether the triangle energy  overcomes the threshold $sJ_\infty$ or not. 
Using a standard separation of scales (see \cite{BG2}) we set 
$$
\II_\eps=\left\{T\in \TTeps^c (\Omega)\,:\, \sum_{k=1}^3 J (|\langle \nabla u_\eps, \eta^k\rangle|)> sJ_\infty\right\}
$$
and define $v_\eps\in SBV(\Omega,\R^2)$ as 
\begin{equation}
v_\eps=\begin{cases}u_\eps & \hbox{ if }\,T\in\TTeps\setminus\II_\eps\\
I & \hbox{ if }\,T\in\II_\eps\\
\end{cases}	
	\label{veps}
\end{equation}
with $I$ denoting the identity deformation. 
By construction the sequence $(v_\eps)$ lies in $SBV (\Omega, \R^2)$ and for any $\eps>0$ its jump set $S_{v_\eps}$ 
is contained in the boundary of the union of the triangles in $\II_\eps$. 

Moreover, for any open set $B$ compactly supported in $\Omega$ inequality \eqref{stima1} can be rewritten in terms of $v_\eps$ as 
\begin{equation} \label{stimaveps}
\eeps (u_\eps)\ge \dfrac{2}{\sqrt{3}\eps}\int_B\hat J (\nabla v_\eps)\dx+csJ_\infty\eps \#(\II_\eps) 
\end{equation}
for a positive constant $c$. 
This implies at once that the functions $u_\eps$ and $v_\eps$ differ in a set of vanishing Lebesgue measure so that it is enough to prove 
that $v_\eps$ is compact in $SBV (\Omega, \R^2)$. 

Thanks to hypothesis ${\rm (iii)}$ on $J$ the set 
$$
\mathcal K=\{A\in\mathcal M(2\times 2;\R)\,:\, \hat J(A)\le s\,J_\infty\}
$$ 
is a compact set in $M(2\times 2;\R)$ and this easily provides the estimate 
$$
\sup_\eps\|\nabla v_\eps\|_{L^2(B,\R^2)}<+\infty. 
$$
On the other hand hypothesis ${\rm (iv)}$ ensures that there exists a constant $c=c(s)$ such that \eqref{stimaveps} can be further sharpened as 
\begin{equation} \label{stimaveps2}
\eeps (u_\eps)\ge \dfrac{1}{\eps}\int_B\sum_{k=1}^3  c(s) (|\langle \nabla v_\eps, \eta^k\rangle|-1)^2\dx+c sJ_\infty\Huno(S_{v_\eps}\cap B). 
\end{equation}
Arguing by polar decomposition we also deduce that for any $A\in \mathcal K$ there exists $R=R(A)\in SO(2)$ such that 
$\sum_{k=1}^3 (|\langle A, \eta^k\rangle|-1)^2\ge c |A-R|^2$, thus 
\begin{equation} \label{stimaveps3}
\eeps (u_\eps)\ge \dfrac{c(s)}{\eps}\int_B \dist^2 (\nabla v_\eps, SO(2))\dx+c sJ_\infty\Huno(S_{v_\eps}\cap B). 
\end{equation} 
Taking into account hypothesis \eqref{controlloL1} a straightforward application of Theorem \ref{compactnessgsbv}, 
and the  $L^1$ convergence to $0$ of $v_\eps-u_\eps$, 
yields that there exists $u\in SBV(\Omega,\R^2)$ with $\Huno(S_u)<+\infty$ such that $u_\eps\to u$ in $L^1 (\Om, \R^2)$. 

To prove $\rm (ii)$ we take advantage of a relaxation argument together with some rigidity estimates. 
Indeed, as a first step we prove that $\nabla u (x) \in SO(2)$ for almost every $x\in\Omega$. 
To this end denote $\psi (A):=\dist^2(A,SO(2))$ and estimate the right-hand side of \eqref{stimaveps3} as  
\begin{equation} \label{stimaveps4}
\eps\eeps (u_\eps)\ge c(s)\int_B \psi^{qc} (\nabla v_\eps)\dx+c\eps  sJ_\infty\Huno(S_{v_\eps}\cap B). 
\end{equation} 
Passing to the liminf and using \eqref{lscbulk} we get at once that 
$$
\int_B \psi^{qc} (\nabla u)\dx=0. 
$$
Hence $\psi^{qc} (\nabla u(x))=0$ almost everywhere in $\Omega$ and by applying Lemma \ref{lemma} with $A=\nabla u(x)$ we also get that 
the approximate gradient $\nabla u$ is a rotation for almost every $x\in B$. 
Once this fact is established $\rm (ii)$ follows by applying  Theorem \ref{rigidity} recursively to $u$ with $\Omega$ replaced by any open set $B$ compactly supported in $\Omega$ and then 
letting $B$ invading $\Om$. 
\endproof

\begin{remark}\rm
Hypothesis \eqref{controlloL1} is essential to deduce a compactness result for sequences equibounded in energy 
and avoids the un-physical situation of particles escaping to infinity  (modelled by arbitrarily large translations). 
\end{remark}

The following lemma completes the proof of Proposition \ref{compactness}. This kind of results are widely used in 
variational problems involving crystal microstructures. 
\begin{lemma}\label{lemma}
Let $\psi (A):=\dist^2(A,SO(2))$ and let $A\in \mathcal M(n\times n,\R)$ be such that $\psi^{qc} (A)=0$. Then 
$A\in SO(2)$. 
\end{lemma}
\proof
For $A\in \mathcal M(2\times 2,\R)$ fixed, taking into account Proposition \ref{envelopeform}, let $(\varphi_k)_k$ contained in $W^{1,\infty}_0(\Om,\R^2)$ be such that 
$$
\lim_k\ave_\Om \dist^2\bigl(A+\nabla\varphi_k(x),SO(2)\bigr)\dx=\psi^{qc}(A).  
$$
Setting for almost every $x\in\Om$ $B_k(x)={\rm argmin}\dist (A+\nabla \varphi_k (x), SO(2))$, we have 
$$
\int_\Om|\nabla\varphi_k|^2\dx\le 2\int_\Om |A-B_k(x)|^2\dx + 2\int_\Om \dist^2\bigl(A+\nabla\varphi_k(x),SO(2)\bigr){\rm dx}\le c. 
$$
Hence it is not restrictive to assume that $\varphi_k\rightharpoonup \varphi_\infty$ and $B_k\rightharpoonup B_\infty$ weakly in $W^{1,2}_0(\Om,\R^2)$ and $L^2(\Om,\R^2)$, respectively.  
By lower semicontinuity, using that $\psi^{qc} (A)=0$, we also get 
$$
0=\psi^{qc} (A)=\lim_k\ave_\Om |A+\nabla\varphi_k(x)-B_k(x)|^2\dx\ge \ave_\Om |A+\nabla\varphi_\infty(x)-B_\infty (x)|^2\dx. 
$$
Thus $\psi\left(A+\nabla\varphi_\infty(x)\right)=0$ for almost every $x\in\Om$ and the claim follows from the classical rigidity theorem in $W^{1,2}(\Om,\R^2)$ and the fact that $\varphi\in W^{1,2}_0(\Om,\R^2)$. 
\endproof

\bigskip

\medskip

We underline that in the proof of Proposition \ref{compactness} any choice of the parameter $s\in (0,1)$ ensures the validity of estimate     
\eqref{stimaveps4}. Moreover, assuming that $u_\eps$ converges to a given $u$ in $L^1 (\Om;\R^2)$, any sequence $v_\eps=v_\eps (s)$,  
defined in \eqref{veps}, still converges to $u$ weakly in $SBV(\Omega, \R^2)$ and the energy contribution is proportional to $\Huno (S_{v_\eps})$. 
Since $S_{v_\eps}$ is a sequence of rectifiable sets with {\itshape normal coordinates} this suggests that in the passage to the liminf on $\eps$ 
on the surface part of the right-hand side of \eqref{stimaveps4} we may obtain a lower bound with a surface-type energy 
maintaining the simmetries of the hexagonal lattice. 
In the sequel we built the right surface energy arguing by pairwise interactions along lattice directions 
and exploiting more refined techniques in geometric measure theory.  
\smallskip

Before entering into details we need to introduce some more tools. 
Let $\psi:\R^2\to [0,+\infty)$ be the $1$-homogeneous map such that  
$\{\psi\le 1\}$ coincides with the convex hull of the set $\textsc{S}$ of the unitary vectors of the lattice $L_1$ and let 
$\psi^*:\R^2\to [0,+\infty)$ its dual norm; i.e., $\psi^*$ is the polar function of $\psi$ defined as 
$$
\psi^*(\nu)=\sup_{|\xi|=1}\dfrac{\langle\nu,\xi\rangle}{\psi(\xi)}. 
$$
An easy computation shows that 
$$
\psi^* (\nu)=\sup_{k=1,2,3}|\langle\nu,\eta_k\rangle| 
$$ 
and $\psi^*$ is the $1$-homogeneous functions whose unitary ball is the convex hull of the coordinate directions $\textsc{D}$ scaled by a factor $2/\sqrt{3}$. In addition the following lemma holds true.  
\begin{lemma}\label{dual}
If $\psi^*$ is as above then 
$\displaystyle2\psi^*(\nu)=\sum_{k=1}^3|\langle \nu,\eta_k\rangle|$
for all $\nu\in\R^2$. 
\end{lemma}
\proof 
A direct computation shows that the inequality holds true as an equality for $\nu\in \textsc{D}$. 
Hence one can argue locally in each sector of amplitude $\pi/3$ by using the linearity of $\psi^*$ on such portions of $\R^2$ 
and the result for the coordinate directions. 
The claim follows from the invariance of the inequality under rotations with angle $\pi/3$.  
\endproof

By means of Lemma \ref{dual} in the next proposition we will provide a lower bound on $\Gamma$-$\liminf$ of $\eeps$  by a suitable anisotropic surface energy. Since by Proposition \ref{compactness} the $\Gamma$-$\liminf$ of $\eeps$ 
is finite only on $SBV (\Omega,\R^2)$ we prove the estimate in that functional space.

\begin{proposition}\label{liminfbound} 
Let $\varphi=J_\infty (4/\sqrt{3})\psi^*$, then for any $u\in SBV(\Omega,\R^2)$ it holds 
\begin{equation}\label{optimalbelow}
\Gamma\hbox{-}\liminf_{\eps\to0^+}\eeps (u)\ge \int_{S_u} \varphi(\nu_u)\,d\Huno . 
\end{equation}
\end{proposition}

\proof 
Let $\{u_\eps\}_\eps\subset L^1(\Om;\R^2)$ be fixed with $u_\eps\to u$ in $L^1 (\Om, \R^2)$ and $\|u_\eps\|_{L^\infty(\Omega, \R^2)}+\eeps (u_\eps)\le c$. 
It is not restrictive to assume also that $\liminf_{\eps\to0^+}\eeps (u_\eps)=\lim_{\eps\to0^+}\eeps (u_\eps)$ and $u_\eps \to u$ 
almost everywhere in $\Omega$. Note that by Proposition \ref{compactness} $u$ is a piecewise rigid deformation with $\Huno (S_u)<+\infty$. 

We will proceed by a slicing technique, performed only in the lattice directions $\eta_1, \eta_2, \eta_3$, in order to obtain 
the estimate 
\begin{equation}\label{finale}
\lim_{\eps\to0^+}\eeps (u_\eps)\ge \sum_{k=1}^3\dfrac{2}{\sqrt{3}}\int_{S_u}|\langle \nu_u(x),\eta_k\rangle|\dHuno (x) .
\end{equation}

To this end we observe that we can split the energies $\eeps$ by accounting separately the contribution of pairs $(\ii,\jj)$ with 
$\jj-\ii=\pm \eta_k\eps$ and letting $k$ varying in $\{1,2,3\}$. Hence, setting for $v\in \admeps$ 
$$
\eeps^k(v)=\sum_{ \ii\in L_\eps\cap\Omega}\eps J\Big(\Big|\dfrac{v(\ii)-v(\ii+\eps\eta_k)}{\eps}\Big|\Big),  
$$
we get 
$$
\eeps (u_\eps)=\eeps^1(u_\eps)+\eeps^2(u_\eps)+\eeps^3 (u_\eps).  
$$
Thus it is enough to prove that for any $k$
\begin{equation}\label{slicingk}
\lim_{\eps\to0^+}\eeps^1(u_\eps)\ge \dfrac{2}{\sqrt{3}}\int_{S_u}|\langle \nu_u(x),\eta_k\rangle| \,\dHuno (x). 
\end{equation}
As the lattices $L_\eps$ and so the energies appearing in both sides of \eqref{slicingk} are invariant under rotations of $\pi/3$ 
we confine our attention to prove \eqref{slicingk} for $k=1$ and $\eta_1=e_1$. 

Note that the functionals $\eeps^1$ consist of a superposition of $1$-dimens\-ional discrete energies related to the sublattices 
$\eps\Z\times \{m\sqrt{3}/2\}$ for $m$ even varying in $\Z$ and to the lattices $\eps(\Z+(1/2))\times \{m\sqrt{3}/2\}$ for $m$ odd. 
Hence, as a first step we will use the usual separation of scales argument on these $1$-dimensional discrete energies in order to rewrite them 
in a suitable $1$-dimensional integral form and then we glue the information back to the $2$-dimensional setting by a slicing  procedure.   
More precisely, for any $m\in \Z$ set 
$$
S_m=\R\times [m\sqrt{3}/2,(m+1)\sqrt{3}/2).
$$ 
Note that the stripes $S_m\cap\Omega$ give a partition of $\Omega $. For any $s\in (0,1)$  
we will construct a sequence $\{w_\eps^s\}\subset SBV(\Omega,\R^2)$  with $1$ dimensional profile along the direction $e_1$ and depending only in the first variable in each stripe.  
Thus let $s\in (0,1)$ be fixed and set 
$$
I_\eps =\{\ii\in L_\eps\cap\Omega\,:\, J(|u(\ii)-u(\ii+\eps e_1)|/\eps)\ge s J_\infty \}
$$  
(for the sake of notation we drop the dependence on $s$ in what follows).  

For any $m\in \Z$ define $w_\eps$ on $\R\times\{m\sqrt{3}/2\}$ to be equal to the value $u_\eps (\ii)$ on $(\ii,\ii+\eps)\times \{m\sqrt{3}/2\}$ if 
$\ii\in I_\eps$ and to be the affine interpolation of the values $u_\eps(\ii), u_\eps (\ii+\eps e_1)$ on any other interval  $(\ii,\ii+\eps)\times \{m\sqrt{3}/2\}$. 
Eventually, extend $w_\eps$ on $S_m\cap\Omega$ as $w_\eps(x_1,x_2)=w_\eps(x_1, m\sqrt{3}/2)$. 

Arguing as in the proof of Proposition \ref{compactness} we infer that 
\begin{equation}\label{stimaweps}
\int_\Omega |\nabla w_\eps|^2\dx +\eps \# (I_\eps)\le c 
\end{equation} 
and that $w_\eps\to u$ in $L^1 (\Omega,\R^2)$ and, up to subsequences, also almost everywhere in $\Omega$. 
In addition 
\begin{equation}\label{stimaeeps1}
\eeps^1(u_\eps)\ge sJ_\infty \eps \# (I_\eps). 
\end{equation}
Taking into account that 
$$\# (I_\eps)=\sum_{m\in\Z} \#(I_\eps\cap S_m)$$ 
and also that 
$$\#(I_\eps\cap S_m) =\#(S(w_\eps^{e_1,y}))
$$ 
for all $y\in S_m\cap \Pi^{e_1}$,
\eqref{stimaeeps1} can be rewritten as  
\begin{equation}\label{stimaeeps2}
\eeps^1(u_\eps)\ge sJ_\infty  \dfrac{2}{\sqrt{3}}\int_{\Pi^{e_1}}\#(S(w_\eps^{e_1,y}))\dHuno(y). 
\end{equation}
By passing to the liminf in both sides of \eqref{stimaeeps2} and by applying Fatou's Lemma we get 
\begin{equation}\label{stimaeeps3}
\lim_{\eps\to0^+}\eeps^1(u_\eps)\ge sJ_\infty  \dfrac{2}{\sqrt{3}}\int_{\Pi^{e_1}}\liminf_{\eps\to0^+}\#(S(w_\eps^{e_1,y}))\dHuno(y). 
\end{equation}
On the other hand, estimate \eqref{stimaweps} ensures that for $\Huno$-almost every $y\in \Pi^{e_1}$ the sequence $w_\eps^{e_1,y}$ is precompact 
in $SBV (\Omega^{e_1,y}, \R^2)$ and converges in measure to $u^{e_1,y}$.  
Hence by the $1$-dimensional analogue of the lower semicontinuity Theorem \ref{semicontinuity} we infer that 
\begin{equation}\label{stimaeeps4}
\lim_{\eps\to0^+}\eeps^1(u_\eps)\ge sJ_\infty  \dfrac{2}{\sqrt{3}}\int_{\Pi^{e_1}}\#(S(u^{e_1,y}))\dHuno(y). 
\end{equation}
Moreover, thanks to Theorem \ref{slicing}(a) with $g=1$, we have 
\begin{equation}
\int_{\Pi^{e_1}}\#(S(u^{e_1,y}))\dHuno(y)= \int_{S_u} |\langle \nu_u (x), e_1\rangle|\dHuno(x).
\end{equation}
Letting $s\to 1$ concludes the proof of claim \eqref{finale}. Eventually, Lemma \ref{dual} yields \eqref{optimalbelow}. 
\endproof

\section{Upper estimates for a class of ``opening cracks"}\label{opaopa}
In this section we show the optimality of the bound \eqref{optimalbelow} provided in Proposition \ref{liminfbound} for 
a class of functions with `opening cracks' with respect to the anisotropies inherited by the lattice. 

In the next section more complex geometries will be taken into account and the occurrence of different phenomena will be highlighted. 
Actually, the representation of the limit energy seems to take into account several factors, not all of local nature. 

In order to clarify the parameters playing a role in the asymptotic behaviour of $\eeps$, we prefer to deal with the case where only two rotations $R^+,R^-$ are involved in the target deformation $u$ first. 
In this setting the crack $S_u$ splits $\Omega$ into two regions $\Om^+,\Om^-$, in general not connected, each one underlying a rigid motion. 
We will show that in this case the $\Gamma$-limsup of $\eeps$ is finite even if the request of orientation-preserving  
recovery sequences affects substantially the form of the limit energy. 

In this process a relevant condition that translates  the positive-deter\-min\-ant constraint through the crack is the following:
\begin{equation}
\langle u^+(x)-u^-(x), R^\pm\nu\rangle\ge 0. 
	\label{incompen0}
\end{equation} 
In fact, the inequality  above ensures that any small triangle crossing the fracture site 
maintains positive (non negative) area in the codomain. 
Actually, due to the discrete environment, condition \eqref{incompen0} has to be assumed for $\nu\in \textsc{D}$, since 
the only triangles entering in the construction have sides parallel to the directions lying in $\textsc{S}$.

We start by proving the following characterization of the $\Gamma$-limsup whenever $S_u$ is contained in a straight line and \eqref{incompen0} holds. 

\begin{proposition}\label{upestimate}
For $R^{\pm}\in\sodue$ and $\bar x,\nu,q^\pm\in\R^2$ with $\nu\in \textsc{D}$, let $u$ be defined as 
\begin{equation}
u(x)=\begin{cases}R^+x+q^+ & \hbox{ if }\,\langle x-\bar x,\nu\rangle > 0,\ x\in\Omega\\
R^-x+q^- & \hbox{ if }\,\langle x-\bar x,\nu\rangle \le 0,\  x\in\Omega.\\
\end{cases}	
	\label{ubasic}
\end{equation}
Assume that $R^+\ne-R^-$ and for $\Huno$-almost every $x\in S_u$ $R^\pm,q^\pm,\nu$ satisfy the condition 
\begin{equation}
\langle u^+(x)-u^-(x), R^\pm\nu\rangle\ge 0
	\label{incompen}
\end{equation}
then 
\begin{equation}\label{limsup1}
\Gamma\hbox{-}\limsup_{\eps\to0^+} \eeps (u)\le \int_{S_u}\varphi (\nu)\dHuno=2 J_\infty\Huno(S_u).
\end{equation}
\end{proposition}
\proof 

We claim that thanks to the hypothesis $R^+\ne-R^-$ we may assume the stronger separation hypothesis 
\begin{equation}\label{positivity}
\langle u^+(x)-u^-(x), R^\pm\nu\rangle\ge\delta> 0
\end{equation}
for some fixed $\delta>0$ and for $\Huno$-almost every $x\in S_u$. 

Indeed, if \eqref{positivity} does not hold, it is enough to choose a vector $v$ such that $\langle v,R^\pm\nu\rangle>0$ (set for instance $v=R^+\nu+R^-\nu$) 
and consider the sequence $u_\delta$ defined replacing $q^+$ with $q^++\delta v$ in \eqref{ubasic}. 
Clearly $u_\delta\to u$ in $L^2(\Om,\R^2)$ and the left-hand side term of \eqref{limsup1} is continuous along such a sequence. 
Condition \eqref{positivity} plays a key role in showing that the sequence $(u_\eps)$ defined as the pointwise interpolation of $u$ in the nodes of the lattice $L_\eps\cap\Om$ is admissible and accounts for the desired $\Gamma$-limsup estimate. 
 
 Thus, set $u_\eps(\ii)=u(\ii)$ for any $\ii\in L_\eps\cap\Omega$. Clearly $u_\eps\to u$ in $L^1(\Om,\R^2)$; it remains to check that 
 the positive-determinant constraint is satisfied in any triangle contained in $\Omega$. 

\begin{figure}[h!]
\centerline{\includegraphics [width=5.2in]{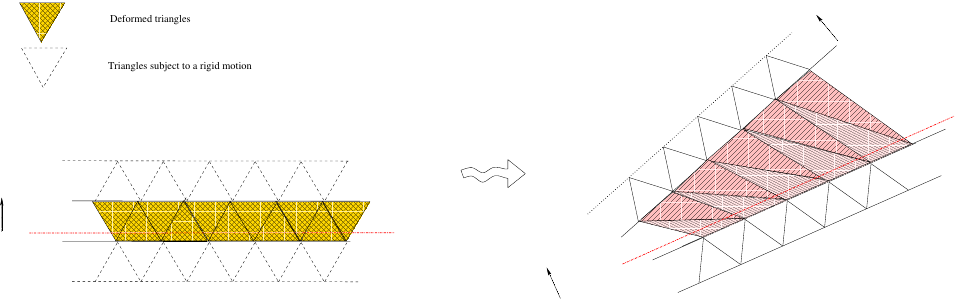}}
\caption{(a) Triangles deformed by the pointwise interpolation} \label{limsup}
\end{figure}

Before proceeding with the computation we  label the triangles `crossing the fracture site'.   
By rotational and translational invariance we may assume $\nu=(0,1)$ so that $S_u=\{x\in \Omega\,:\,x={\bar x}_2\}$.  
Let $N_\eps:=[2{\bar x}_2/\sqrt{3}\eps]$ and set 
$$\deps=\{m\in\Z\,: (m\eps, \sqrt{3}N_\eps/2)\in L_\eps\cap\Omega\}$$
 and 
$$\ii_m=(m\eps, \sqrt{3}N_\eps\eps/2)\qquad \hbox{ for }m\in \deps.
$$ 
Note that $u$ coincides with a `positive' rotation on the vertices of those triangles  not contained in the strip  
$$
\mathcal S= \{x\,:\, \sqrt{3}N_\eps\eps/2\le x_2\le \sqrt{3}(N_\eps+1)\eps/2\}.
$$ 
Hence, in order to ensure that  $u_\eps\in\admeps$, it suffices to check condition \eqref{incompen} 
for triangles $\textsc{T}$ with vertices respectively  $\ii_m,\ii_m+\eps\eta^1,\ii_m+\eps\eta^2$ and $\ii_m,\ii_m+\eps\eta^2,\ii_m+\eps(\eta^2-\eta^1)$ 
(see Fig. 2). This leads to prove the following inequalities:
$$
\langle (u^-(\ii_m+\eps\eta^1)-u^-(\ii_m))^\perp, u^+(\ii_m+\eps\eta^2)-u^-(\ii_m)\rangle\ge 0,
$$
$$ 
\langle (u^+(\ii_m+\eps\eta^2)-u^+(\ii_m+\eps(\eta^2-\eta^1))^\perp, u^+(\ii_m+\eps\eta^2)-u^-(\ii_m)\rangle\ge 0
$$
for $m\in\deps$. 
These can be compactly rewritten as 
\begin{equation}\label{deter}
\eps \langle R^\pm\nu, u^+(\ii_m+\eps\eta^2)-u^-(\ii_m)\rangle\ge 0. 
\end{equation}
We claim that these conditions are fulfilled for $\eps$ small enough. Indeed, for any infinite collection of indices $\{m_\eps\}\subset \Z$ with $m_\eps\in\deps$, up to subsequences, we may assume that 
$\ii_{m_\eps}\to \hat x$ as $\eps\to 0+$. As a consequence $u^\pm(\ii_{m_\eps}+\eps\eta^1)\to u^\pm(\hat x)$ and 
 $u^\pm(\ii_{m_\eps})\to u^\pm(\hat x)$, with 
 ${\hat x}_2={\bar x}_2$. 
 Thus, assuming that \eqref{deter} are violated for such a sequence of indices $m_\eps$ will lead to a contradiction to \eqref{positivity} when passing to the limit as $\eps\to 0+$.
 
 Eventually we compute the asymptotic value of $\eeps (u_\eps)$ to show that $(u_\eps)$ is a recovery sequence for the surface energy $\int_{S_u} \varphi(\nu_u)\,d\Huno$. 
 Clearly, the energy contribution in $\eeps (u_\eps)$ reduces to those pairs of indices one of which of type $\ii_m$ with $m\in\deps$; i.e., 
\begin{eqnarray}
 \notag\eeps(u_\eps)&=&\sum_{m\in \deps}\eps J\Big(\dfrac{|u^+(\ii_m+\eps\eta^2)-u^-(\ii_m)|}{\eps}\Big)\\
\notag & &+\sum_{m\in \deps}\eps 
 J\Big(\dfrac{|u^+(\ii_m+\eps(\eta^2-\eta^1))-u^-(\ii_m)|}{\eps}\Big)\\ \nonumber
 &= & 2 \sum_{m\in \deps}\eps (J_\infty+o(1)) \\
 &=& 2\eps J_\infty\#\{m\in \deps: \ii_m\in S_u\} +o(1).\label{ult}
\end{eqnarray}
Passing to the limit as $\eps\to 0$ and plugging the equality 
 $$
 \lim_{\eps\to0^+} \#\{m\in \deps: \ii_m\in S_u\}=\Huno (S_u)
 $$
  in \eqref{ult}  the conclusion is achieved.
\endproof
%
%



The next proposition generalizes the result of Proposition \ref{upestimate} in the case when only two rotations are involved in the deformation and $S_u$ consists of segments having normals in $\textsc{D}$.     

\begin{proposition}\label{upestpoli}
Let $R^{\pm}\in\sodue$ and $q^\pm\in\R^2$ be given with $R^+\ne-R^-$. Let $S$ be a connected union of segments with normal belonging to $\textsc{D}$ and  let $$u(x)=(R^+x+q^+)\chi_{\Om^+}(x)+(R^-x+q^-)\chi_{\Om^-}(x)$$ 
where $\Om^+,\Om^-$ is a Caccioppoli partition subordinated to $S$. 
Assume that for $\Huno$-almost every $x\in S$ it holds 
\begin{equation}\label{openpoli}
\langle u^+(x)-u^-(x), R^\pm\nu (x)\rangle\ge 0, 
\end{equation}
then 
\begin{equation}\label{limsup2}
\Gamma\hbox{-}\limsup_{\eps\to0^+} \eeps (u)\le \int_{S}\varphi (\nu (x))\dHuno=2 J_\infty\Huno(S_u).
\end{equation}
\end{proposition}
\proof 
Up to an approximation argument we may assume that $S=\cup_{k\in {\mathcal I}} S_k$ with $S_k=[i^k,j^k]$ for $i^k, j^k\in L_\eps\cap\Omega$, having constant normal $\nu_k$ lying in $\textsc{D}$ and satisfying $j^k=i^{k+1}$ for any $k$. Thanks to the connectness hypothesis and to condition \eqref{openpoli}, arguing by perturbation with infinitesimal translations, condition \eqref{openpoli} can be assumed to hold in the stronger form $\langle u^+(x)-u^-(x), R^\pm\nu (x)\rangle\ge \delta$, for a positive $\delta$. 
We claim that a recovery sequence is provided defining $u_\eps (\ii)=u^+(\ii)$ for $\ii\in L_\eps\cap \Omega$. Indeed, for any triangle $T\in \TTeps^c$ intersecting $S$ there exists a unique $k\in {\mathcal I}$ such that either $T\cap S$ is a single point in $S_k$ or $T\cap S= [\ii, \ii\pm \eps \nu_k^\perp]$ for some $\ii \in L_\eps\cap \Omega$. In both cases one may perform the same computation as in Proposition \ref{upestimate}, and deduce that $u_\eps$ satisfies the positive-determinant constraint. Eventually a direct computation shows that $\lim_{\eps\to0^+} \eeps (u_\eps ) =J_\infty \sum_{k\in {\mathcal I}} \varphi (\nu_k) \Huno (S_k)$. 
\endproof

The previous approach can be pushed further to obtain the optimality of the bound \eqref{optimalbelow} also for $S_u$ consisting of a line with normal $\nu\not\in \textsc{D}$. In this case we need to impose the opening crack condition \eqref{incompen} in a stronger sense; i.e., \eqref{incompen} must be satisfied also along the two directions $\nu_1, \nu_2\in \textsc{D}$ generating $\nu$ in one of the simplex of the Wulff shape $\{\varphi \le 1\}$. 

\begin{proposition}\label{normalenu}
For $R^{\pm}\in\sodue$ and $\bar x,\nu,q^\pm\in\R^2$, let $u$ be defined as 
\begin{equation}
u(x)=\begin{cases}R^+x+q^+ & \hbox{ if }\,\langle x-\bar x,\nu\rangle > 0,\ x\in\Omega\\
R^-x+q^- & \hbox{ if }\,\langle x-\bar x,\nu\rangle \le 0,\  x\in\Omega.\\
\end{cases}	
	\label{ubasic4}
\end{equation}
Let $\nu_1, \nu_2\in \textsc{D}$ be such that  $\nu/\varphi (\nu)=\lambda \nu_1+(1-\lambda) \nu_2$ for $\lambda\in (0,1)$. 
Assume that for $\Huno$-almost every $x\in S_u$ $R^\pm,q^\pm,\nu$ satisfy the conditions
\begin{equation}
\langle u^+(x)-u^-(x), R^\pm\nu_i\rangle\ge \delta 
	\label{incompen2}
\end{equation}
for $i=1,2$ and for some $\delta >0$. 
Then 
\begin{equation}\label{3}
\Gamma\hbox{-} \limsup_{\eps\to0^+} \eeps (u)\le \int_{S_u}\varphi (\nu)\dHuno.
\end{equation}
\end{proposition}
\proof 

The claim will be proved by an approximation argument, exploiting Proposition \ref{upestpoli} and the lower semicontinuity of the $\Gamma$-limsup. It is not restrictive to assume $S_u$ connected and with endpoints on $\partial \Omega$. 
Note that hypothesis \eqref{incompen2} ensures that $R^+\ne -R^-$. 
Let $h>0$ be a fixed small parameter and select $\{S_h\}_h$ a connected union of segments with normal equal to $\nu_1$ or to $\nu_2$ such that  
\begin{equation}\label{conv}
\lim_{h\to +\infty} \int_{S_h}\varphi(\nu_h(x))\dHuno=\int_{S_u}\varphi(\nu)\dHuno
\end{equation}
and $\lim_{h\to +\infty}\dH (S_h, S_u)=0$. By simply extending the polygonals above we may also assume that each $S_h$ has endpoints on $\partial \Omega$ and splits $\Omega$ in two  components $\Omega_h^+, \Omega_h^-$ such that 
$$\lim_{h\to +\infty}|A_h^+\triangle\{x\in \Omega\,:\, \langle x-\bar x,\nu\rangle > 0\}|=0.
$$ 
For $h\in\N$ let $u_h$ be the piecewise rigid function defined as 
$$
u_h(x)=(R^+x+q^+)\chi_{A_h^+}+(R^-x+q^-)\chi_{A_h^-}.
$$ 
It is easily checked that $u_h\to u$ in $L^1(\Omega;\R^2)$ and, by continuity,  $u_h$ satisfies hypothesis \eqref{incompen2} with a smaller positive $\delta$ and $S_u$ replaced by $S_{u_h}=S_h$.    
Applying Proposition \ref{upestpoli} to each $u_h$ we get 
$$\Gamma\hbox{-}\limsup_{\eps\to0^+} \eeps (u_h)\le \int_{S_{u_h}}\varphi (\nu_h)\dHuno .
$$
The conclusion follows taking  property \eqref{conv} into account. 
\endproof

In the following we consider the case in which the jump set $S_u$ consists in a triple point with coordinate normals. 

\begin{proposition}\label{tripunto}
Let $x_0\in \Omega$ be fixed and set 
\[
\Omega_1:=\{x\in \Omega\,:\, \langle x-x_0, {\eta_1}^\perp\rangle \ge 0,\, \langle x-x_0, {\eta_2}^\perp\rangle \ge 0\},
\]
\[
 \Omega_2:=\{x\in \Omega\,:\, \langle x-x_0, {\eta_2}^\perp\rangle \le 0,\, \langle x-x_0, {\eta_3}^\perp\rangle \le 0\},
\] 
\[\Omega_3:=\{x\in \Omega\,:\, \langle x-x_0, {\eta_1}^\perp\rangle \le 0, \, \langle x-x_0, {\eta_3}^\perp\rangle \ge 0\}.
\]
Let $u$ be defined as 
$u(x)=\sum_{k=1}^3 u^k(x) \chi_{\Omega_k}(x)$ with $u^k(x)=R^k x+q^k$ for suitable $R^1, R^2,R^3\in\sodue$ and $q^1,q^2,q^3\in\R^2$. Assume  
that $u$ satisfies  the `opening crack' conditions along $S_u=\bigcup_k (\partial \Omega_k\cap \Omega)$: 
\begin{align}\label{opentripunto}
\langle u^1(x)-u^3(x), R^1{\eta_1}^\perp\rangle >0,  \ 	\langle u^1(x)-u^3(x), R^3{\eta_1}^\perp\rangle >0, \hbox{ on }\Omega_1\cap \Omega_3 \notag \\
\langle u^1(x)-u^2(x), R^2{\eta_2}^\perp\rangle >0,  \ 	\langle u^1(x)-u^2(x), R^1{\eta_2}^\perp\rangle >0,\hbox{ on }\Omega_1\cap \Omega_2 \notag\\
\langle u^3(x)-u^2(x), R^3{\eta_3}^\perp\rangle >0,  \ 	\langle u^3(x)-u^2(x), R^2{\eta_3}^\perp\rangle >0,	\hbox{ on } \Omega_3\cap\Omega_2 
\end{align}
and that $u$ satisfies the further compatibility condition 
\begin{equation}\label{postripunto}
\langle u^1(x_0)-u^3(x_0), { (u^2(x_0)-u^3(x_0)})^\perp\rangle >0 
\end{equation}
 on the triple point $x_0$. Then 
\begin{equation}\label{4}
\Gamma\hbox{-}\limsup_{\eps\to0^+} \eeps (u)\le \int_{S_u}\varphi (\nu)\dHuno.
\end{equation}
\end{proposition}
\proof
Up to composing $u$ with infinitesimal translations we may assume that, for any $\eps>0$ small enough, the points $x_0-\eps/3(\eta_1+\eta_2), x_0+\eps/3(\eta_3+\eta_1),x_0+\eps/3(\eta_2-\eta_3) $ lie on the lattice $L_\eps\cap \Omega$. Set $u_\eps(\ii)=u(\ii)$ for any $\ii\in L_\eps\cap\Omega$. Clearly $u_\eps\to u$ in $L^1(\Om,\R^2)$; it remains to check that the positive-determinant constraint is satisfied in any triangle intersecting $S_u$. Thanks to hypothesis \eqref{postripunto} we have that there exists $\delta >0$ such that all the scalar products in \eqref{opentripunto} are greater than $\delta$. By performing the same computation as in the proof of Proposition \ref{upestimate} we may deduce the positive-determinant constraint to be  satisfied on any triangle intersecting $S_u$ and not containing $x_0$. The only triangle left aside is then the one having vertices $x_0+\eps/2\sqrt{3}(\eta_2-\eta_1), x_0+\eps/2\sqrt{3}(\eta_3-\eta_1),x_0+\eps/2\sqrt{3}(\eta_2+\eta_3) $. A direct computation shows that 
\begin{eqnarray*}
& &\langle u^1(x_0+\eps/3(\eta_2-\eta_3))-u^3(x_0-\eps/3(\eta_1+\eta_2)),\cr & & (u^2(x_0+\eps/3(\eta_1+\eta_3))-u^3(x_0-\eps/3(\eta_1+\eta_2)))^\perp\rangle =\cr
& &\langle u^1(x_0)-u^3(x_0), { (u^2(x_0)-u^3(x_0)})^\perp\rangle +o(1).
\end{eqnarray*}
Hence the conclusion follows by hypothesis \eqref{postripunto}. 
\endproof

As a conclusion of this section we note that we have proved that the $\Gamma$-limit of $E_\e$
is described by the anisotropic fracture energy
\begin{equation}\label{Griffe}
{\cal F}(u)=\int_{S_u}\varphi(\nu)d\Huno
\end{equation}
 on all $u$ which are piecewise rigid deformations
such that $S_u$ consists of a finite number of lines meeting at triple points and 
the opening-crack conditions in Proposition \ref{tripunto} are satisfied. This description 
extends by continuity to all piecewise rigid deformations $u$ that can be approximated
in energy by sequences of piecewise rigid deformations $u_h$ satisfying such conditions.

\begin{figure}[htbp]
\begin{center}
\includegraphics[width=.5\textwidth]{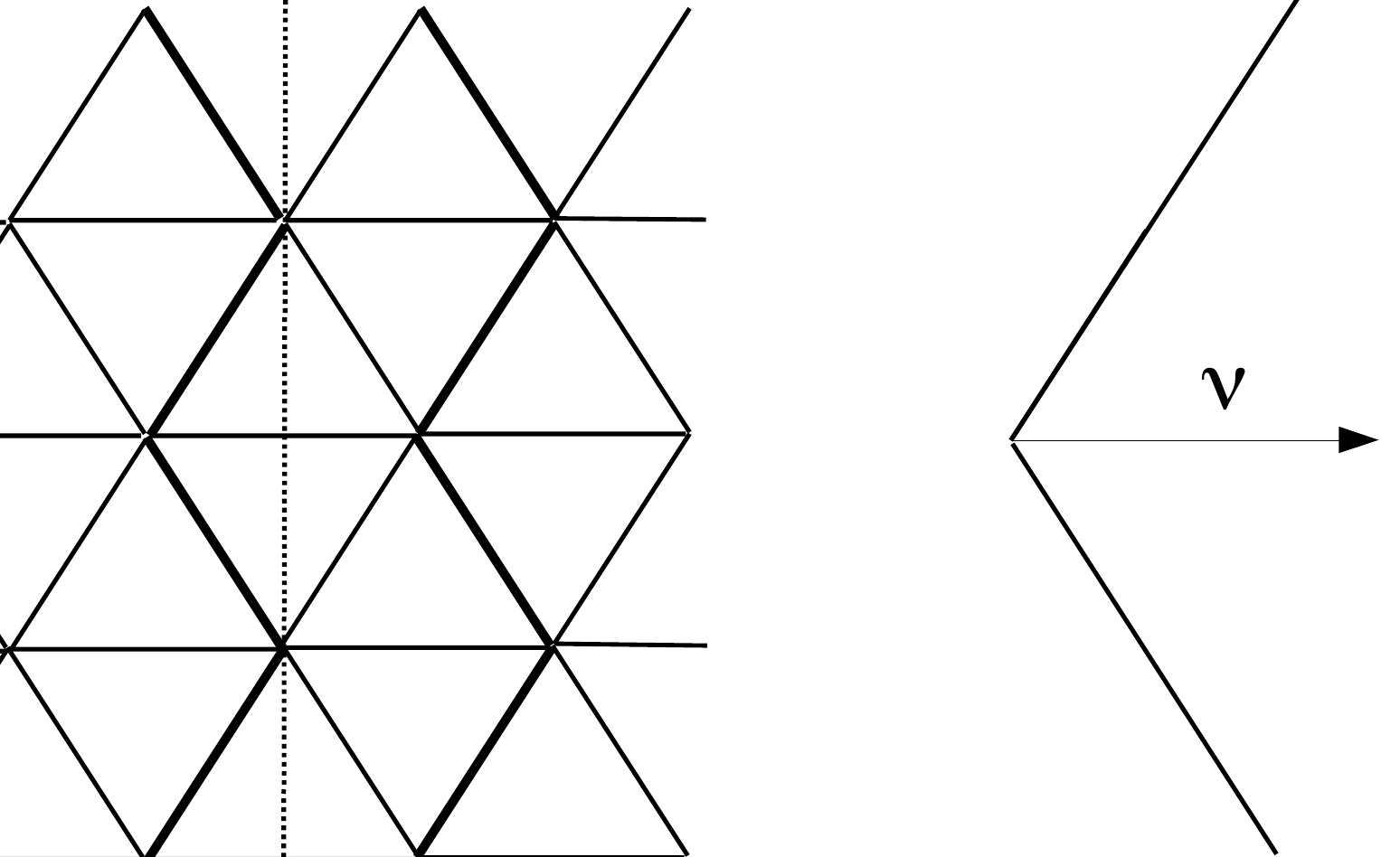}
\caption{A layer corresponding to fracture in the reference configuration, and the corresponding admissibility angle for $u^+-u^-$}
\label{surrel}
\end{center}
\end{figure}
\section{Macroscopical failure of the positive-de\-ter\-mi\-nant constraint}

In the previous sections we have computed the $\Gamma$-limit of our discrete system when the macroscopic configuration satisfies compatibility conditions on the fracture site and at meeting points of three (or more) fracture sites. Those conditions can be regarded as a {\em positive-determinant constraint on the fracture}. 
In this section we will see how all those conditions can be removed. In this way macroscopic configurations ``with negative-determinant fracture'' can be obtained from atomistic configurations satisfying a microscopic positive-determinant constraint, at the expense of a strictly greater energy.

\begin{figure}[htbp]
\begin{center}
\includegraphics[width=.3\textwidth]{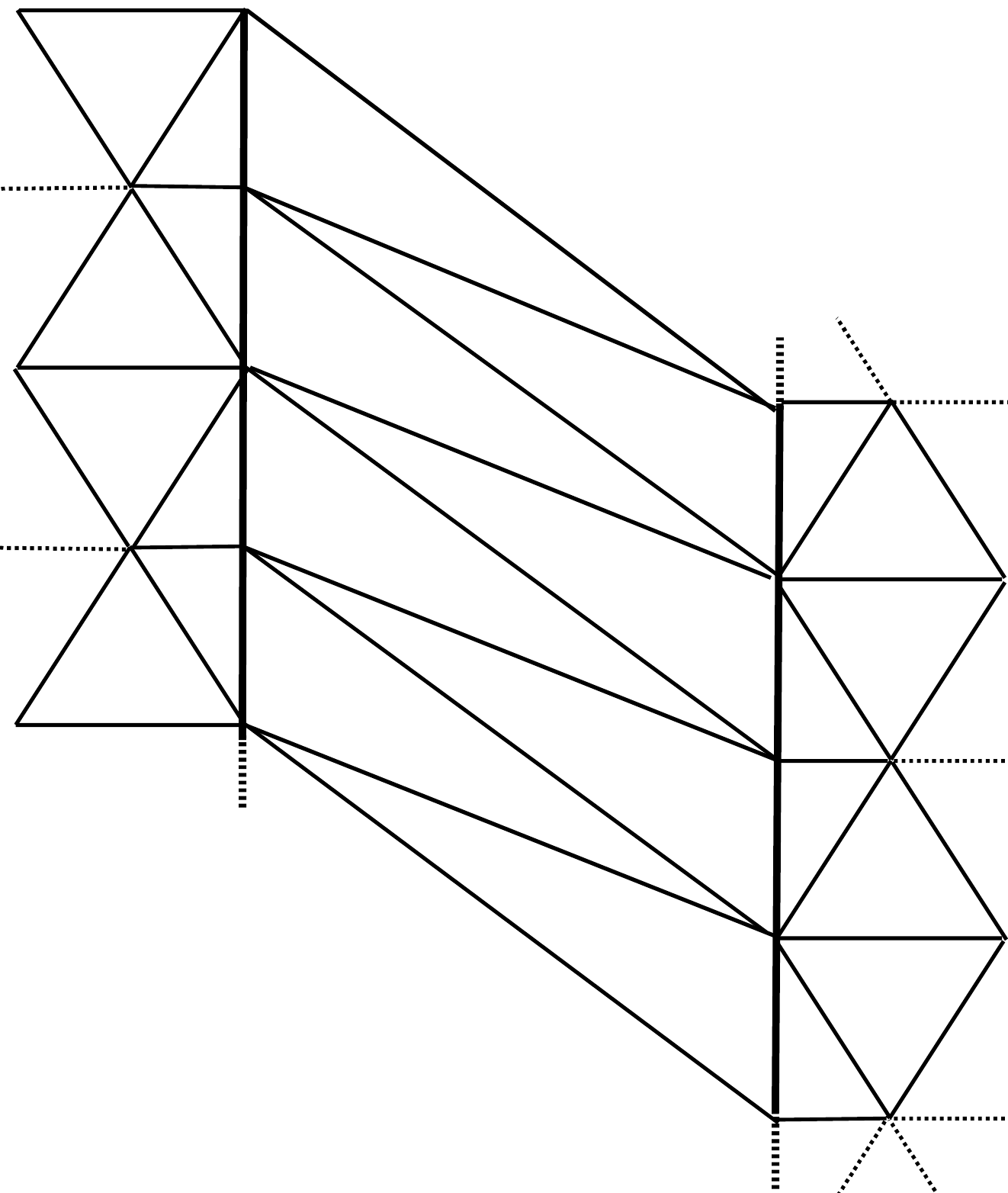}
\caption{Surface relaxation allowing for  $\langle u^+-u^-,\nu\rangle\ge 0$ (deformed configuration)}
\label{surrel2}
\end{center}
\end{figure}
\subsection{Removal of compatibility conditions on the fracture site -- surface relaxation}\label{sure}
The surface energy \eqref{Griffe} has been obtained under a condition that implies that the fracture can be achieved by highly deforming only a single layer of positive-determinant triangles, keeping all other triangles at the minimum of the energy. 
We now show how, still using a single layer of triangles to create the fracture, we may relax conditions (\ref{incompen2}) in the case when $\nu\not\in D$. 

We only illustrate the surface relaxation with an example. Consider a linear crack surface with normal $\nu=(1,0)$, which does not belong to $D$. In Fig.~\ref{surrel} we highlight a layer of triangles that may be used to obtain a fracture satisfying (\ref{incompen2}). If $R^+=R^-=Id$ then the possible values for $q^+-q^-$ lie in the convex angle pictured on the right of the figure.
We may obtain any $q^+-q^-$ with $\langle q^+-q^-,\nu\rangle\ge 0$ if we also deform the triangles neighbouring the `fracture layer'. A possible deformation is given in Fig.~\ref{surrel2},
where each such neighbouring triangle has one side compressed by a factor $\sqrt 3/2$ and
one side compressed by a factor $1/2$. 
This construction gives an estimate for the energy density on the fracture with
$$
2+2 J\Bigl({\sqrt 3\over 2}\Bigr)+ J\Bigl({1\over 2}\Bigr)
$$
(while $\varphi(\nu)=2$).

Note that such a wide relaxation is not necessary for fixed $q^+$, $q^-$ if 
$\langle q^+-q^-,\nu\rangle> 0$. In general then, even in this simple geometry,
we will obtain a surface energy density depending on $u^+-u^-$. 

\begin{figure}
\begin{center}
\includegraphics[width=.35\textwidth]{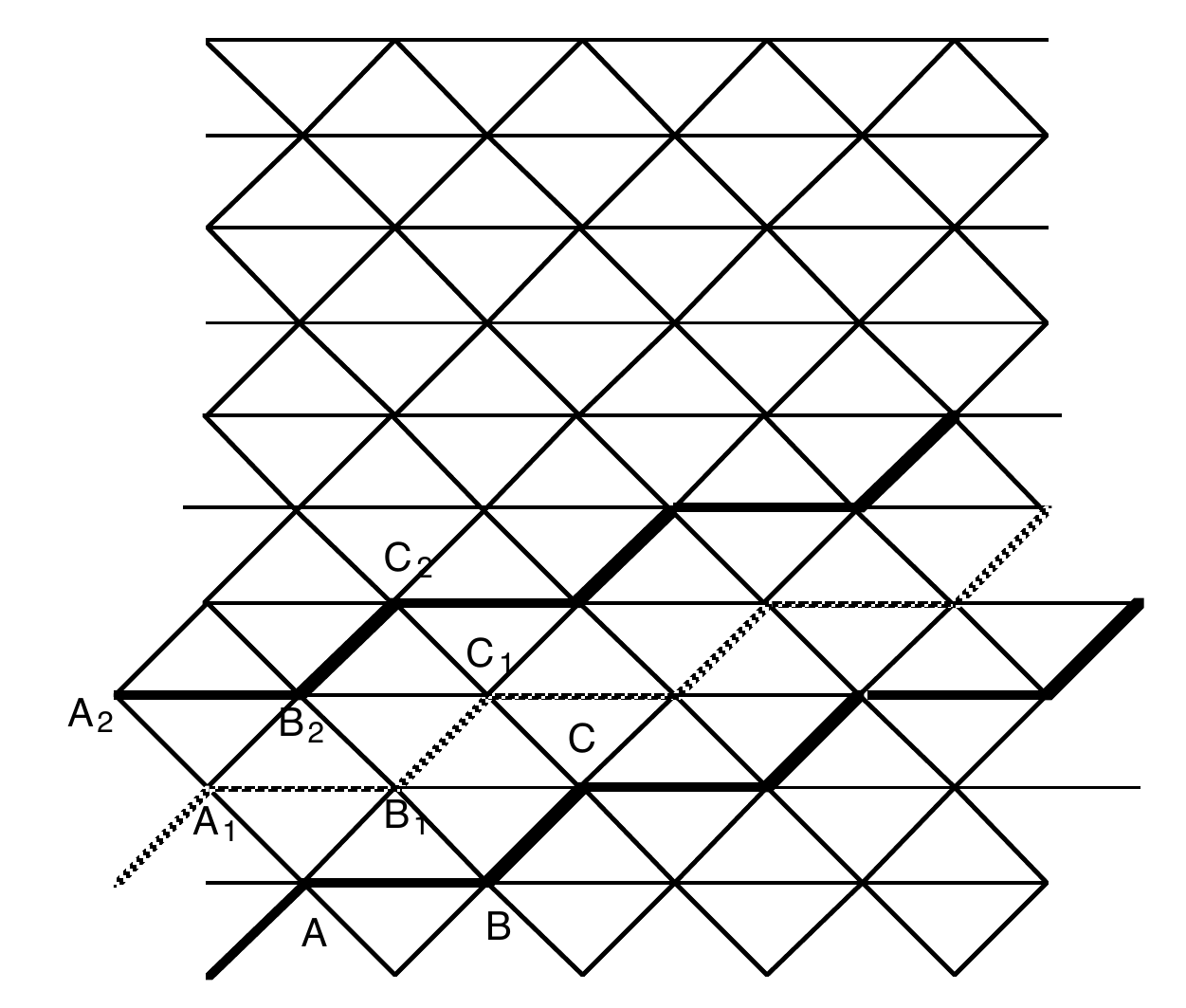}
\includegraphics[width=.55\textwidth]{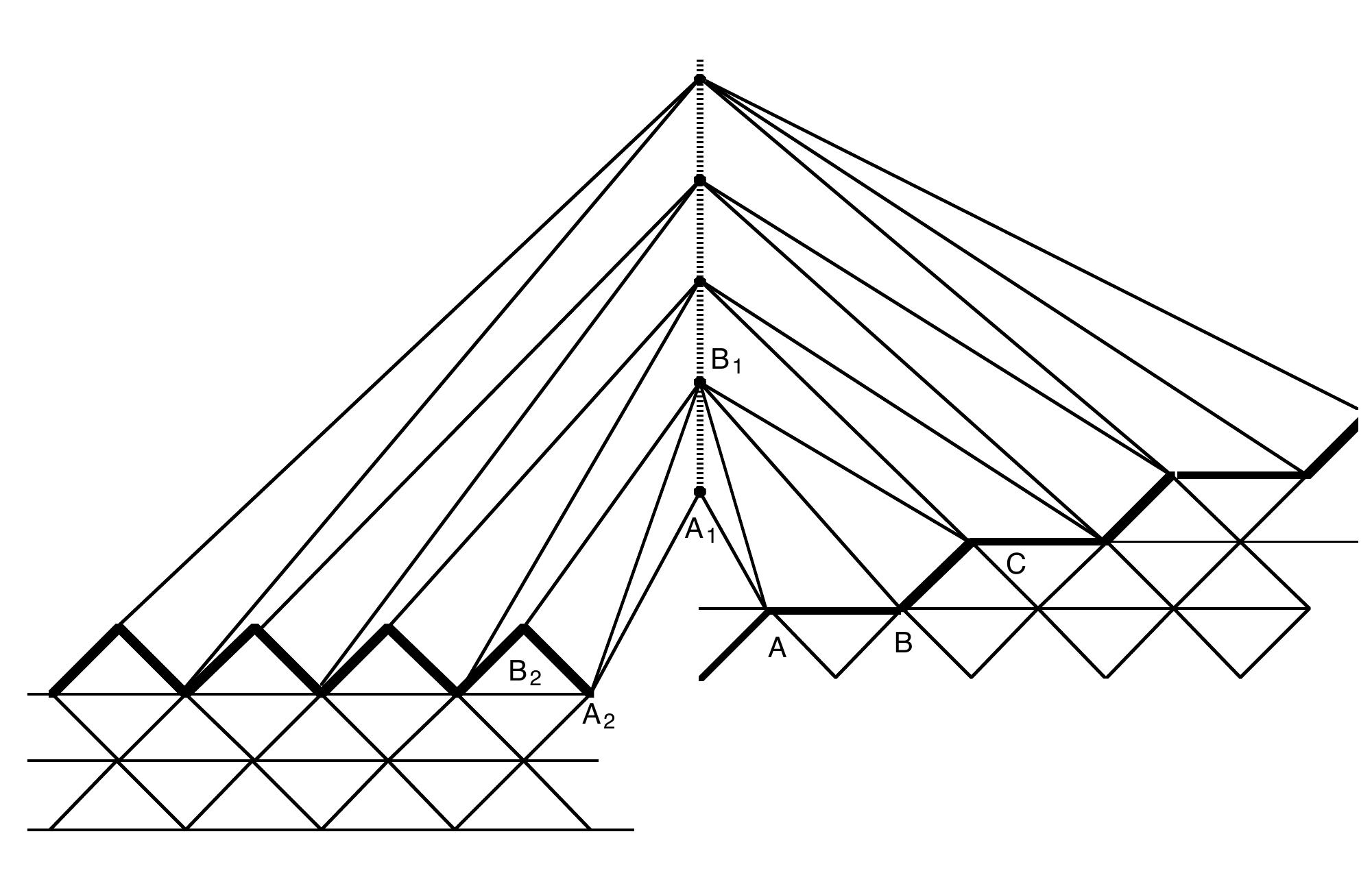}
\caption{Multiple microscopic fracture}
\label{multifrac1-2}
\end{center}
\end{figure}

\begin{figure}[htbp]
\begin{center}
\includegraphics[width=.3\textwidth]{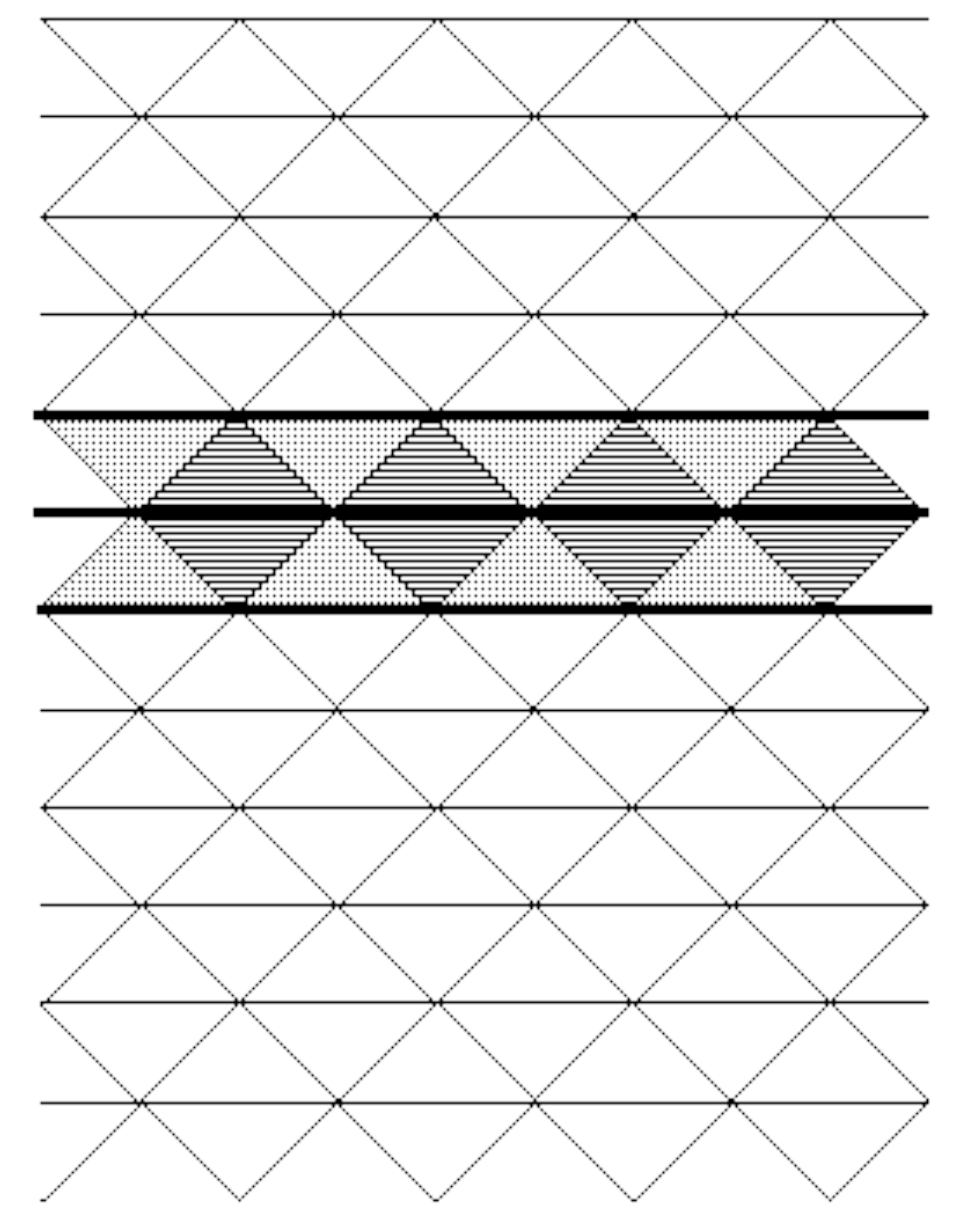}
\includegraphics[width=.5\textwidth]{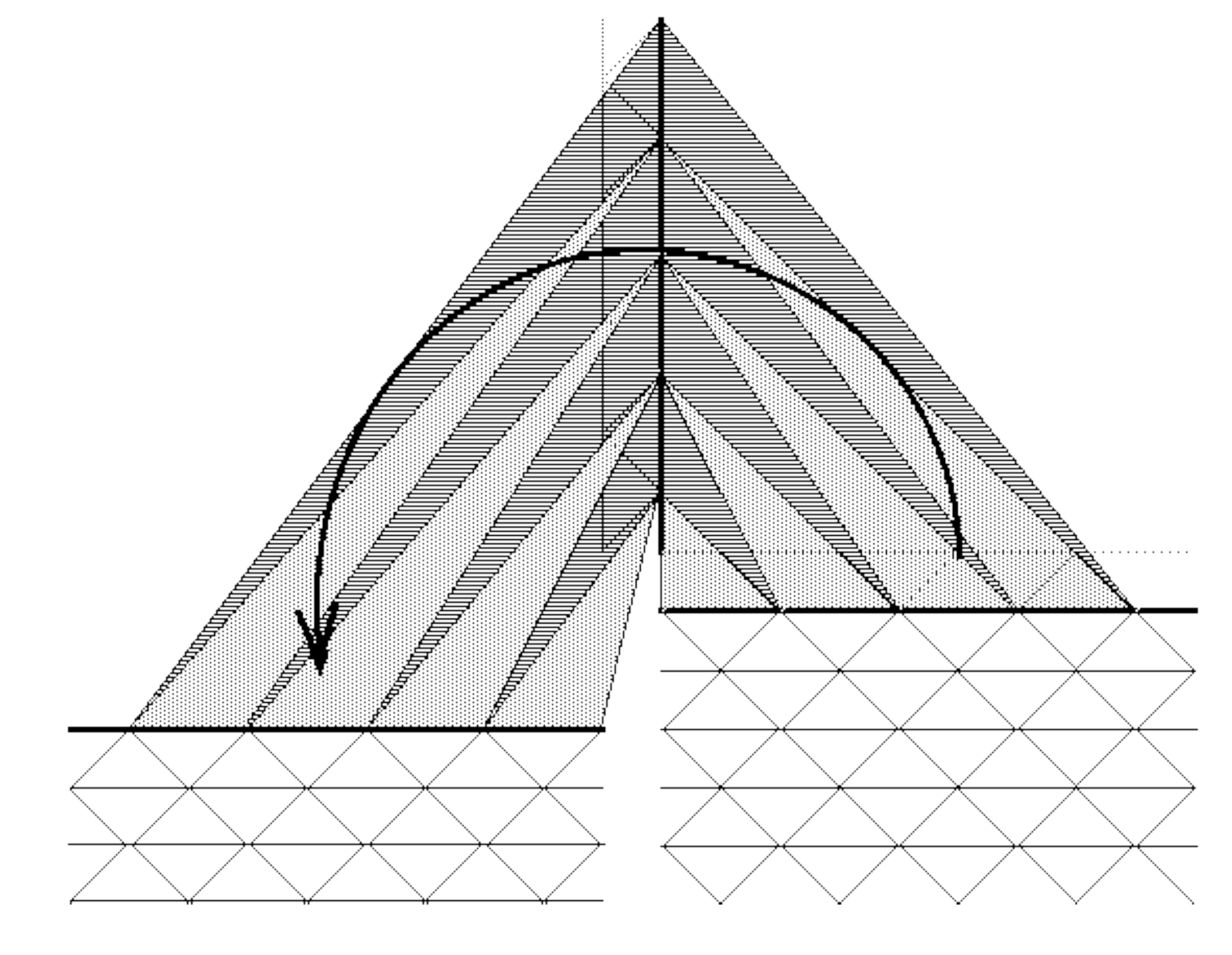}
\caption{Multiple microscopic fracture - reference and deformed configurations}
\label{multifrac3-4}
\end{center}
\end{figure}
\subsection{Removal of compatibility conditions on the fracture site -- multiple microfracture}\label{removal1}
By introducing more layers we may remove all constraints on $S_u$, while the macroscopic energy varies by a factor proportional to the number of additional layers. 

In Figure \ref{multifrac1-2} the macroscopic deformation on the fracture does not satisfy the conditions 
$$\langle u^+(x)-u^-(x), R^\pm\nu_i\rangle\ge 0$$
for the two ``microscopic'' coordinate normals $\nu_i$. However, the discrete functions in the construction depicted in Figure \ref{multifrac1-2} all have positive-determinant interpolations thanks to the introduction of a ``fictitious'' layer of atoms.
In this case the limit energy per length doubles.

\begin{figure}[htbp]
\begin{center}
\includegraphics[width=0.8\textwidth]{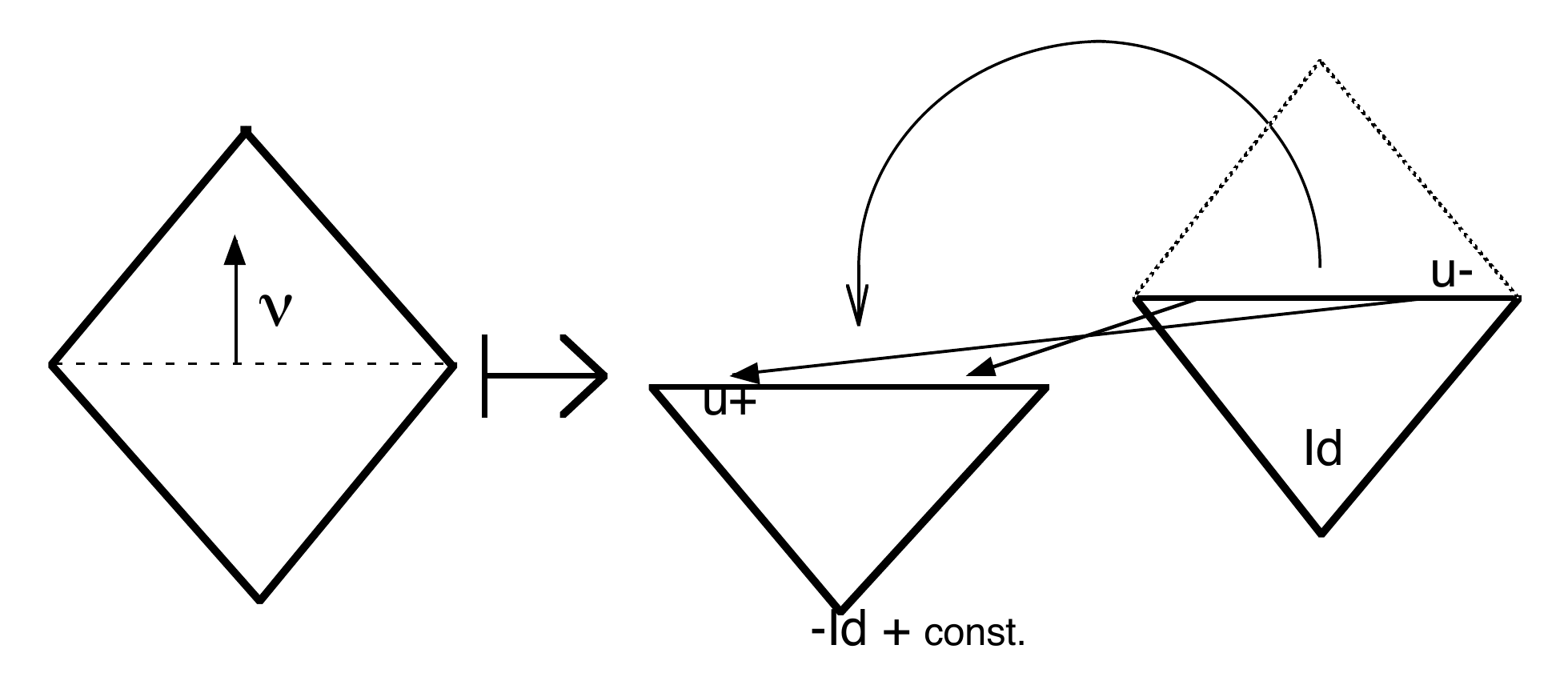}
\caption{Multiple microscopic fracture - macroscopic configurations}
\label{rot-triangle}
\end{center}
\end{figure}

A variation of this example in given in Figures \ref{multifrac3-4} and \ref{rot-triangle} to highlight that the introduction of an extra layer of atomic interactions allows to remove the condition  $\langle u^+(x)-u^-(x), R^\pm\nu\rangle\ge 0$ even when $\nu$ is a coordinate normal. By repeating this process the macroscopic deformation may exhibit a interpenetration phenomenon.

Note that for some deformations surface relaxation may be energetically convenient with respect to multiple-layer fracture. This clearly is the case of the example in Section \ref{sure} when $q^+-q^-$ is close to satisfying the compatibility conditions (\ref{incompen2}).

\begin{figure}[htbp]
\begin{center}
\includegraphics[width=.7\textwidth]{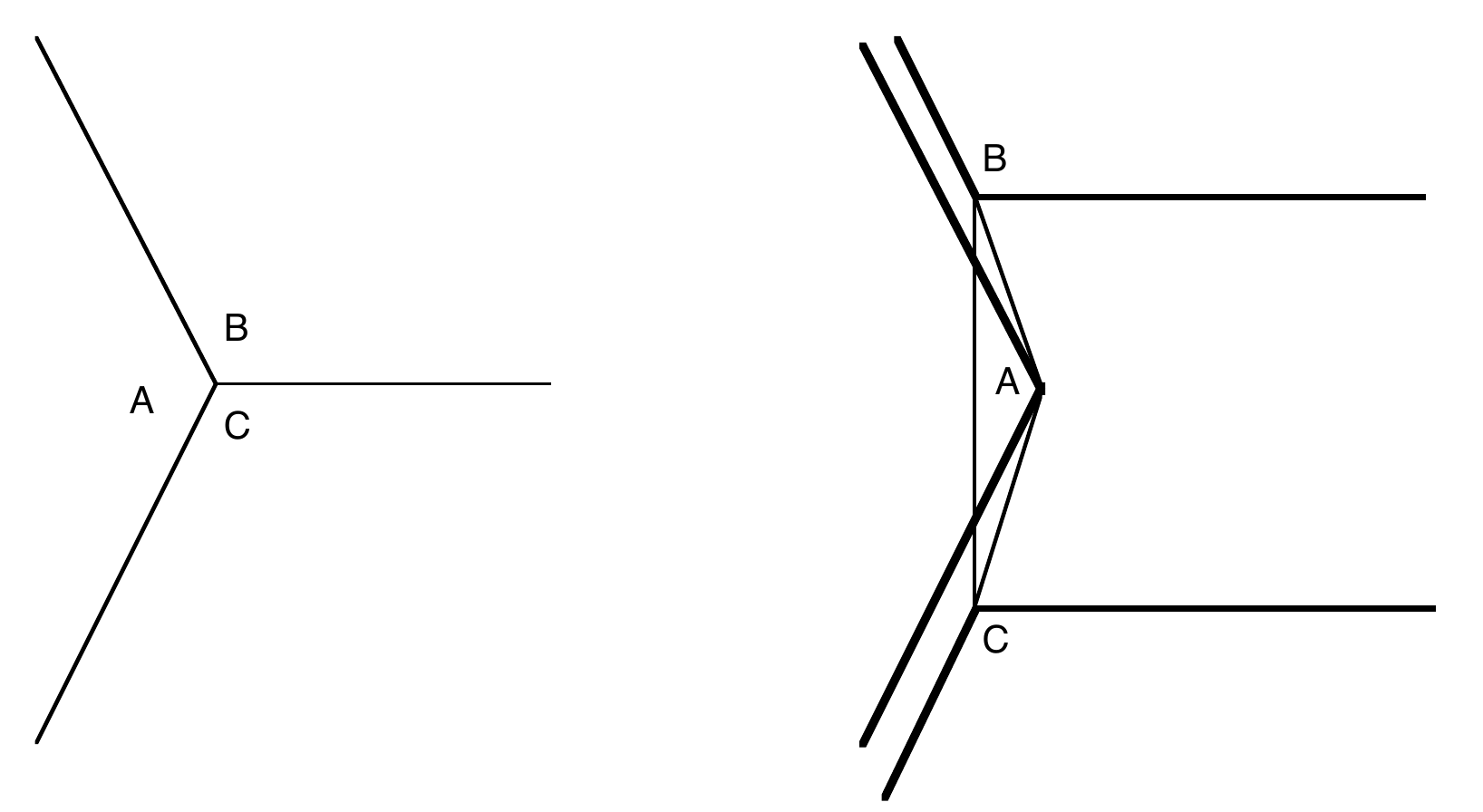}
\caption{Triple point with ``negative determinant'' -- reference and deformed macroscopic deformation}
\label{triplepoint1}
\end{center}
\end{figure}
\subsection{Removal on compatibility conditions on ``triple points'' -- micro-deformed fracture}\label{triplecon}
The condition on triple points in the previous section ensures that the deformation of a microscopic triangle at that point is of positive determinant (and hence, being a single triangle, gives a negligible energy contribution). If such a condition does not hold then the use of (small variations of) pointwise interpolations on the different regions of the underlying partition is not possible, since for $\e$ small there will always be a microscopic triangle whose vertices are mapped in three points which fail the positive-determinant constraint. However, it is possible to use a different interpolation by introducing an additional microscopic fracture enclosing small sets where the deformation is not in $SO(2)$. Note that this is possible if such sets have the dimension of an interface. 

\begin{figure}[htbp]
\begin{center}
\includegraphics[width=.5\textwidth]{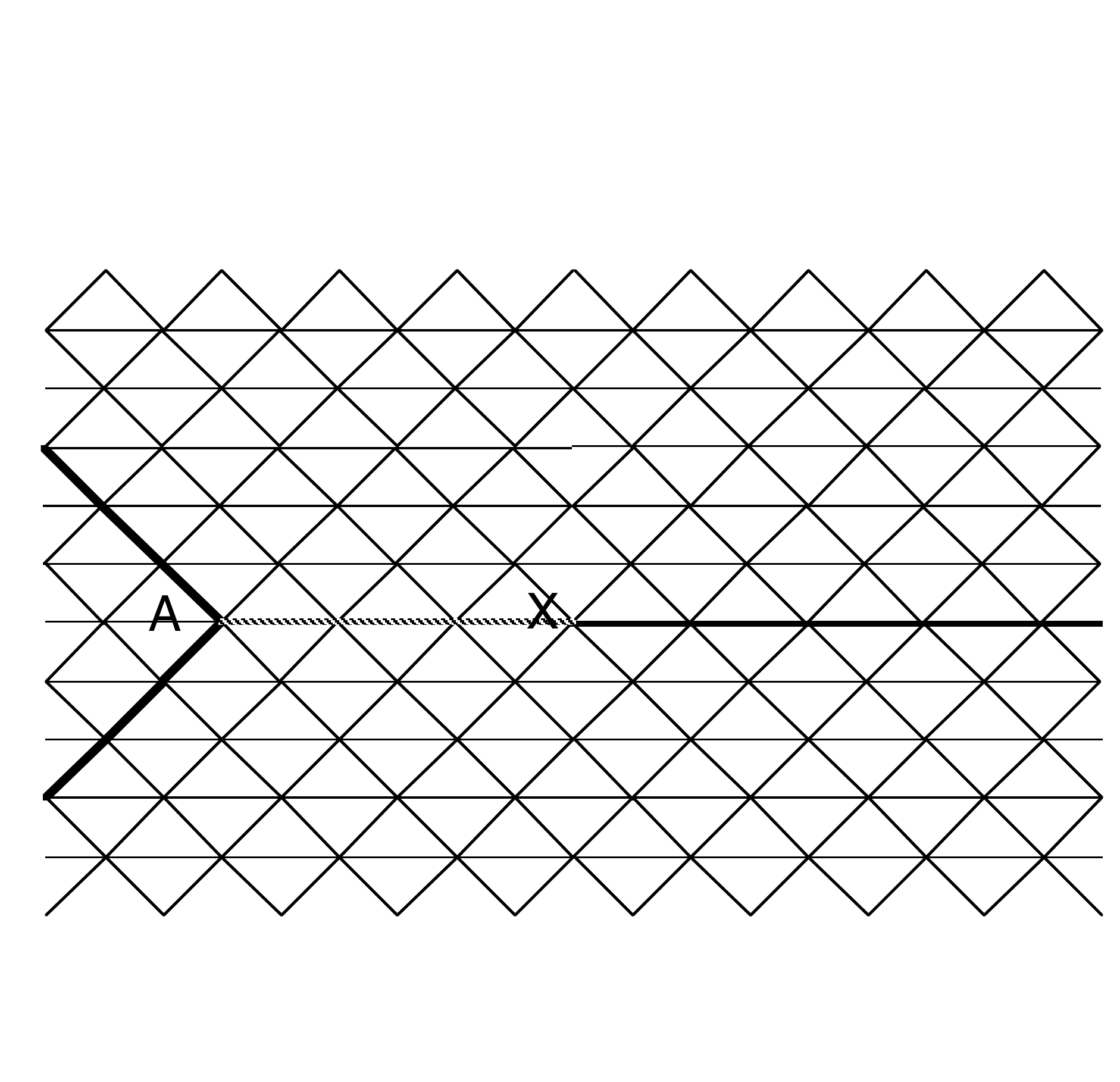}\qquad
\includegraphics[width=.4\textwidth]{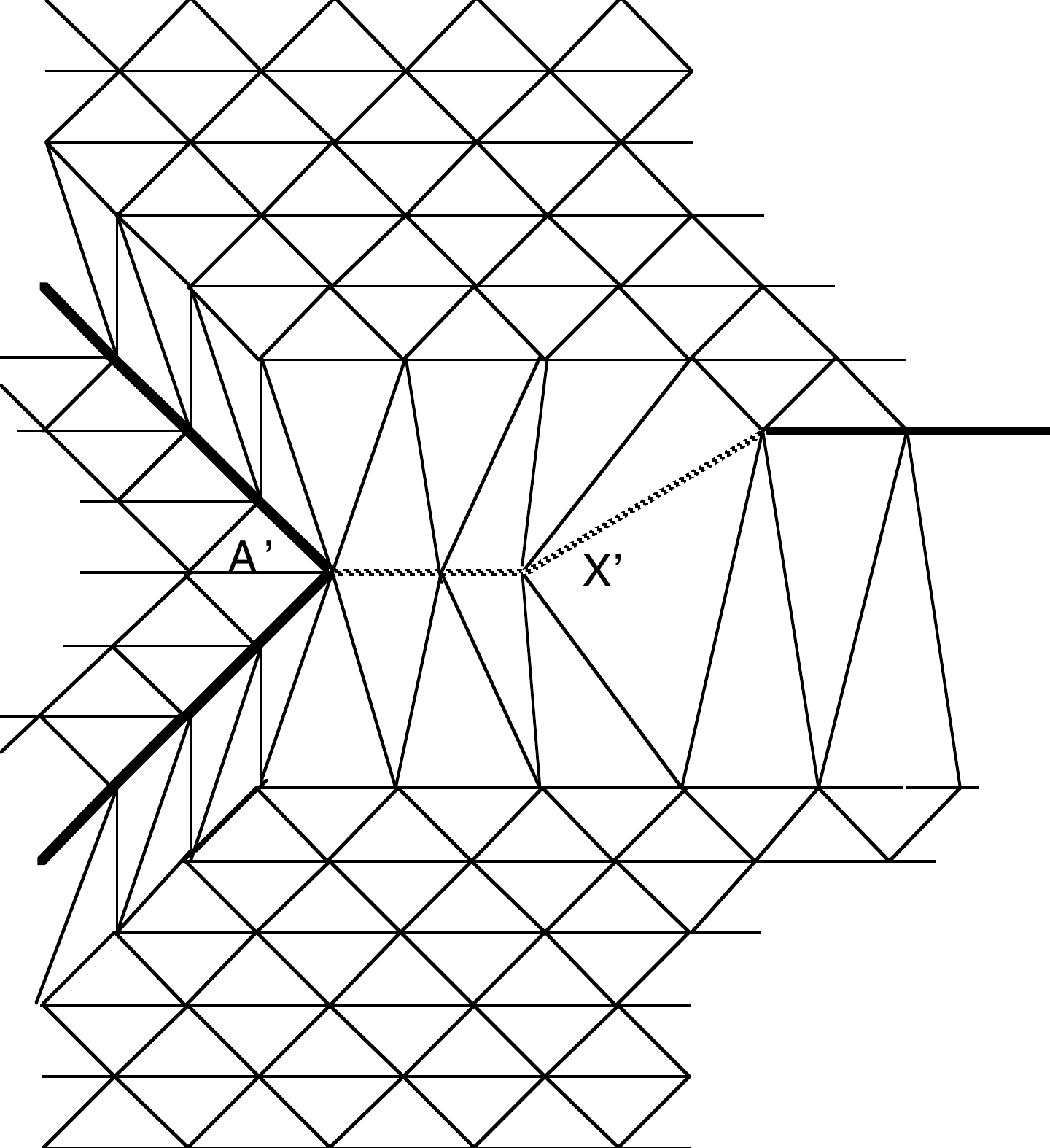}
\caption{Triple point with ``negative determinant'' -- reference and deformed microscopic deformations}
\label{triplepoint2-3}
\end{center}
\end{figure}

In Figure \ref{triplepoint1} it is represented a deformation with a triple point failing the positive-determinant condition.
In this case we introduce a microscopic approximation as represented in Figure \ref{triplepoint2-3}.

On the fracture site we introduce a segment $[A,X]$ where the pointwise single-layer interpolation of the jump is substituted by a double-layer approximation. Note that the image $[A',X']$ of this segment undergoes an additional linear deformation. Note that the energy of this approximation provides the additional contribution
$$
\Bigl(\varphi(\nu)+ J\Bigl({[A',X']\over [A,X]}\Bigr)\Bigr)\Huno([A,X]),
$$
where the second term is due to the compression of the triangles on the segment $[A,X]$ in the reference configuration. 
This energy depends on the choice of $X$ and $X'$, which are variables in the construction (satisfying some constraints due to the positive-determinant requirement in the resulting construction). This shows that even in this simple case an optimization problems arises between the introduction of an additional microfracture and a microscopic compression. Of course, more complex constructions with more parameters can be also introduced.

\subsection{Global failure of impenetrability constraints -- optimal decomposition and healing microfractures}
\begin{figure}[htbp]
\begin{center}
\includegraphics[width=.6\textwidth]{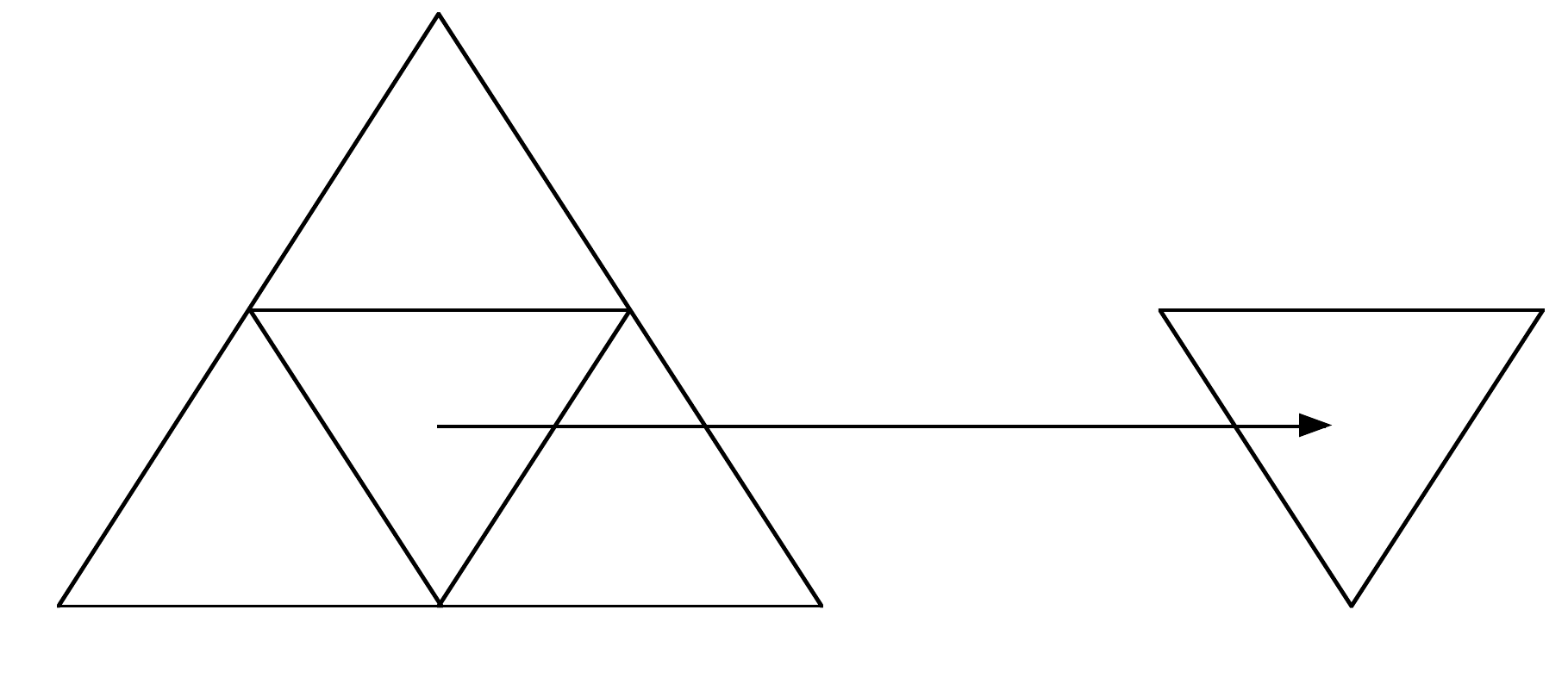}
\caption{Deformation violating the impenetrability constraint}
\label{triangle1}
\end{center}
\end{figure}

In the constructions illustrated above we could exhibit a microscopic recovery sequence by working separately on each fracture site or triple points. In the presence of a complex geometry of the domain, besides a use of those constructions one has also to take into account the possibility of introducing further ``fictitious'' microscopic interfaces to get around 
impenetrability constraints. As a simple example, we may consider the deformation in Figure \ref{triangle1}, 
\begin{figure}[htbp]
\begin{center}
\includegraphics[width=.6\textwidth]{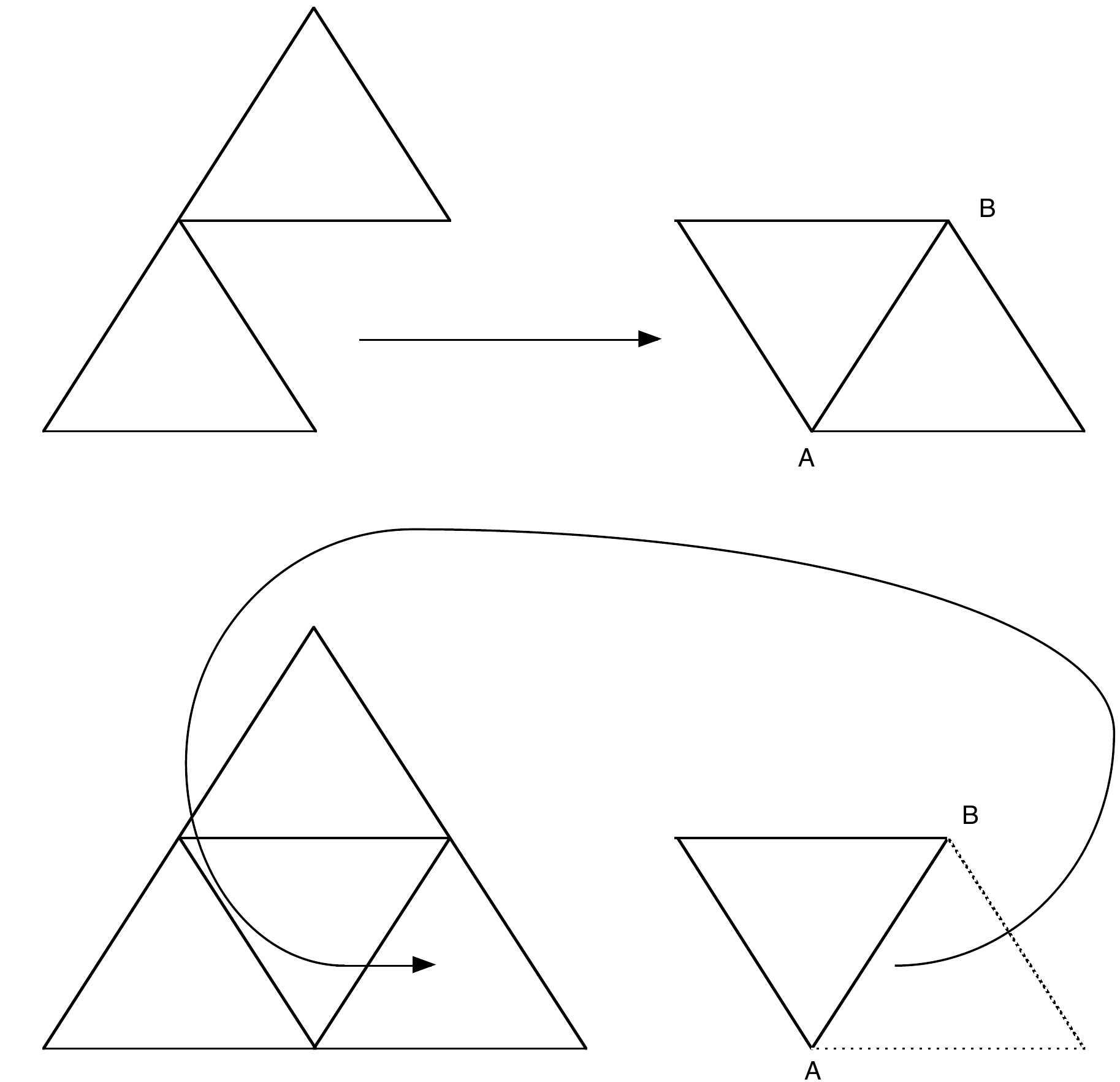}
\caption{Construction of approximations satisfying the positive-determinant constraint}
\label{triangle2}
\end{center}
\end{figure}
where the central smaller triangle is removed and translated from its position in the larger triangle.
This macroscopic deformation can be approximated by microscopic ones all satisfying the positive-determinant constraint (see Figure \ref{triangle2}). One such approximation can be obtained by translating a rhombus and subsequently composing this translation with a deformation rotating half of this rhombus as described in Section \ref{removal1}
(see also Figure \ref{rot-triangle}). Note that this last rotation entails the introduction of one or more microscopic fictitious layers of atoms.

\begin{figure}[htbp]
\begin{center}
\includegraphics[width=.6\textwidth]{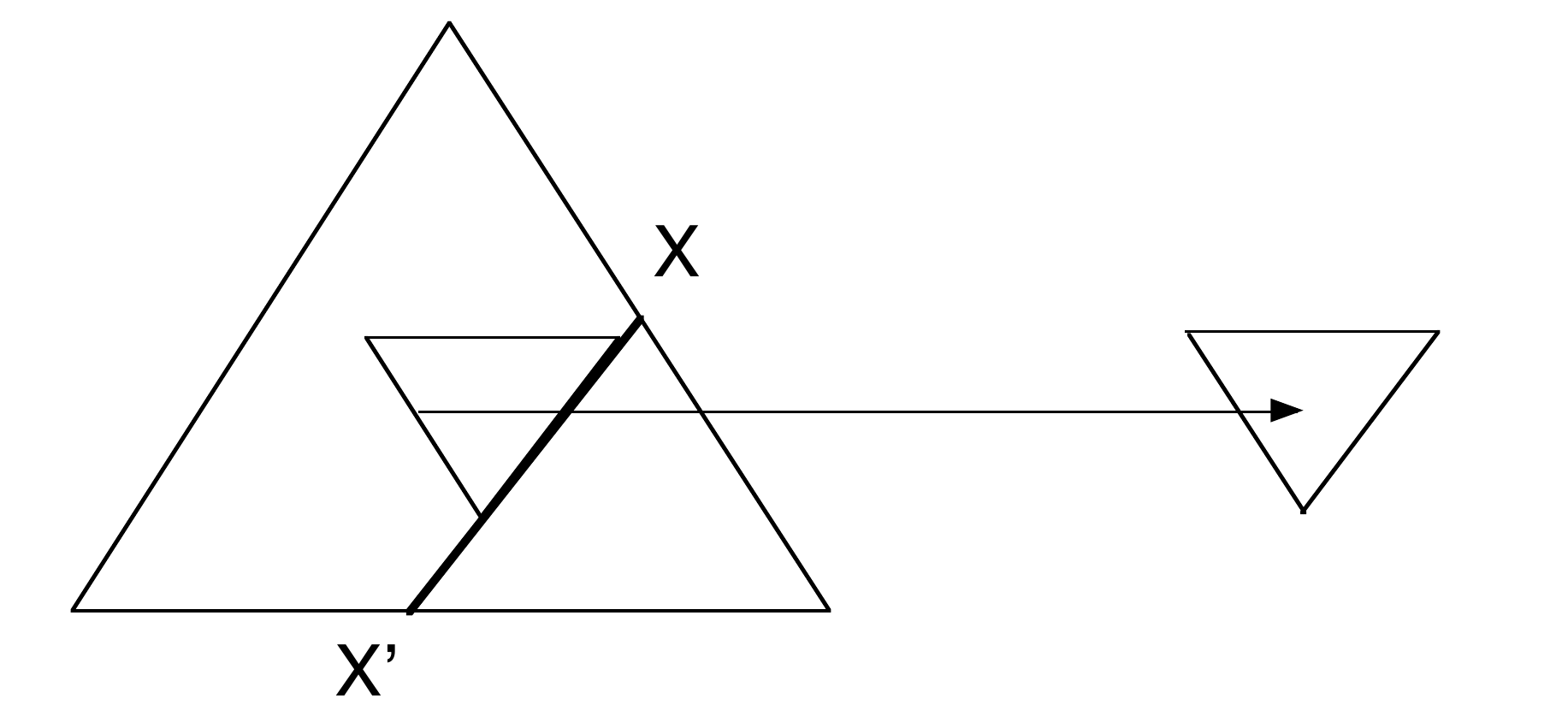}
\caption{Deformation that can be obtained with auxiliary fractures}
\label{triangleinside}
\end{center}
\end{figure}
Another simple example is depicted in Figure \ref{triangleinside}, where the triangle to be removed is strictly contained in the interior of a larger triangle. In this case, in order to proceed as in the previous construction, one has to introduce an ``auxiliary fracture'' as the segment $[X,X']$ in the figure. 
The determination of the optimal shape and location of such auxiliary fracture sites is clearly a complex optimization problem. Note that in this case, also the determinant constraint for triple points has to be taken into account.

\section{Necessary conditions for opening cracks}
In Section \ref{lobaloba} we have shown a lower bound with an anisotropic Griffith fracture energy,
which is optimal on a family of displacements with an opening-crack condition on the fracture site
(Section \ref{opaopa}). We now show conversely that if the limit energy at a point $x_0\in S_u$ is not greater than the lower bound then necessarily the function $u$ satisfies a opening-crack condition.

We now consider discontinuity points where the fracture energy density  
is minimal; i.e., the inequality in \eqref{optimalbelow} is sharp,
and derive necessary conditions on the crack opening. To this end we introduce the measures 
\begin{equation}
\mu_\e=\sum_{ (\ii,\jj)\in \NNeps (\Omega) }\eps J\Big(\Big|\dfrac{u_\e(\ii)-u_\e(\jj)}{\eps}\Big|\Big).
\end{equation}
These measures are nothing else than a way to measure locally the energy $F_\e(u_\e)$.
If $\mu_\e$ is a bounded sequence of measures and $\sup_\eps\|u_\eps\|_{L^\infty (\Omega, \R^2)} < +\infty$, then, arguing as in the proof of Proposition \ref{compactness}, we infer that $u_\e$ is precompact in $SBV(\Omega,\R^2)$ and that $u_\e\to u$ in $L^1$. We also suppose that the weak$^*$ limit $\mu$ of $\mu_\e$ exists.

\newcommand{\LM}[1]{\hbox{\vrule width.2pt \vbox to#1pt{\vfill \hrule
width#1pt
height.2pt}}}
\newcommand{\LL}{{\mathchoice {\>\LM7\>}{\>\LM7\>}{\,\LM5\,}{\,\LM{3.35}\,}}}

\begin{proposition}[necessity of an opening-crack condition]\label{nece}
Let $u_\e\to u$, and let $x_0\in S_u$ be such that
\begin{equation}\label{stiff}
{d\mu\over d{\cal H}^1\LL S_u}(x_0)\le \varphi(\nu_u(x_0)).
\end{equation}
Then we have
\begin{equation}\label{bocia}
\langle u^+(x_0)- u^-(x_0),R^\pm \nu_u(x_0)\rangle \ge 0,
\end{equation}
where $R^\pm\in SO(2)$ are the two constant matrices coinciding with $\nabla u$ on both sides of $S_u$ 
at $x_0$. 
\end{proposition}

\proof
For the sake of brevity we will denote $\nu_0=\nu_u(x_0)$.

Let $Q^{\nu_0}$ be a square centered in $0$, with side length $1$ and 
an edge orthogonal to $\nu_0$.  
With fixed $\rho>0$ and $y\in Q^{\nu_0}$, let 
$$
v_{\e,\rho}(y)= u_\e(x_0+\rho y),
$$
which, by definition of $S_u$ converges as $\e\to0$ and $\rho\to0$ to the function
$$
\widetilde u(y)=\begin{cases}
u^+(x_0) &\hbox{ if $\langle y,\nu_0\rangle \ge 0$}\cr
u^-(x_0) &\hbox{  if $\langle y,\nu_0\rangle < 0$}.
\end{cases}
$$

By the weak$^*$ convergence of $\mu_\e$ and 
\eqref{stiff} we deduce that
$$
E_\e(u_\e,Q^{\nu_0}_\rho(x_0))\le \rho\,\varphi(\nu)+o_\rho(1)+o_\e(1)
$$
where 
$Q^{\nu_0}_\rho(x_0)$ is a cube centered in $x_0$, with side length $\rho$ and 
an edge orthogonal to $\nu_0$,
and $o_\rho(1)$ and $o_\e(1)$ are infinitesimal as $\rho\to 0$ and $\e\to 0$,
respectively.

We fix $s>0$ and set 
$$
{\cal S}^s_\e=\Bigl\{T\in {\cal T}^c_\e(\Omega): \Bigl|{u_\e(\ii)-u_\e(\jj)\over\e}\Bigr|> s \hbox{ for at least 
two sides $(\ii,\jj)$ of }T\Bigr\}.
$$
 We claim that we may connect the two opposite sides of $Q^{\nu_0}_\rho(x_0)$ parallel to $\nu_0$ with a path $\{T_i: i=1,\ldots, M\}$ 
(depending on $\rho$ and $\e$, but we omit such a dependence for the sake of notational simplicity) consisting of triangles such that $T_i$ and $T_{i+1}$ have a common side and $T_i\in {\cal S}^s_\e$ up to a number 
of indices that is $o(\rho/\e)$.
 Indeed, note that for any $T\not\in  {\cal S}^s_\e$ we have
\begin{equation}\label{wecon}
 \Bigl|{u_\e(\ii)-u_\e(\jj)\over\e}\Bigr|\le 2 s \hbox{ for every side $(\ii,\jj)$ of }T.
\end{equation}
If no path as above exists then we may construct $c\rho/\e$ disjoint paths in $L_\e\cap Q^{\nu_0}_\rho(x_0)$;
i.e., sets of indices $\{\ii^j_n: n=0,\ldots,M_j\}$ with $\ii^j_n\in L_\e$ and $\ii^j_n-\ii^j_{n-1}\in \e S$,
such that 
$$
\langle\ii^j_0-x_0,\nu_0\rangle\le-{1\over 2}\rho+\e,\qquad
\langle\ii^j_{M_j}-x_0,\nu_0\rangle\ge{1\over 2}\rho-\e
$$
and 
\begin{equation}\label{trip}
 \Bigl|{u_\e(\ii^j_n)-u_\e(\ii^j_{n-1})\over\e}\Bigr|\le 2 s \hbox{ for every }n
\end{equation}
(for a construction of such paths we refer to the proof of Theorem 4(ii) in \cite{BP}).
Since $v_{\e,\rho}-u^\pm(x_0)$ converge to $0$ close two opposite sides of $Q^{\nu_0}$ it is not restrictive to suppose
that 
$$
\sum_j\e |u^\e(\ii^j_0)-u^-(x_0)|+ \sum_j\e |u^\e(\ii^j_{M_j})-u^+(x_0)|=\rho \,o_\e(1).
$$
By \eqref{trip} we have
\begin{eqnarray}\label{prit}\nonumber
c\rho|u^+(x_0)-u^-(x_0)|&\le&
\sum_j\e\Bigl(|u^+(x_0)-u^\e(\ii^j_{M_j})|\\ \nonumber
&&+|u^\e(\ii^j_{M_j})-u^\e(\ii^j_0)|+|u^\e(\ii^j_0)-u^-(x_0)|\Bigr)
\\ \nonumber
&\le&\rho\, o_\e(1)+\sum_j\sum_{n=1}^{M_j} \e |u^\e(\ii^j_{n})-u^\e(\ii^j_{n-1})|
\\
&\le&\rho\, o_\e(1)+\sum_j\sum_{n=1}^{M_j} \e^2 2s\le \rho\, o_\e(1)+\rho^2 2s.
\end{eqnarray}
Dividing by $\rho$ and letting $\e$ and $\rho$ tend to $0$ we then obtain $|u^+(x_0)-u^-(x_0)|=0$,
which gives a contradiction.

We consider the connected set 
$$
{\bf T}_\e=\bigcup_{i=1}^M T_i.
$$
Note that by the convergence $v_{\e,\rho}\to\widetilde u$, for all $\delta>0$ the set ${\bf T}_\e\cap Q^{\nu_0}_\rho(x_0)$ will be contained in the strip $\{x: |\langle x-x_0,\nu_0\rangle|\le\delta\rho\}$ for $\e$ sufficiently small. 
It is not restrictive to suppose that $Q^{\nu_0}_\rho(x_0)\cap\partial \bf T_\e$ is composed exactly of two polygonal chains $P^\e_-$ and $P^\e_+$. Note that the set 
$$
\{T_i: \partial T_i\cap P^\e_-\neq\emptyset\}
$$
still gives a path with the same properties of $\{T_i\}$. 
We can therefore suppose that this is our original path. 

The contribution of the energy restricted to interactions in ${\bf T}_\e$  gives
\begin{equation}\label{accure}
E_\e(u_\e, Q^{\nu_0}_\rho(x_0)\cap {\bf T}_\e)\ge J(s) \varphi(\nu_0)\rho(1-\delta)
\end{equation}
(see Proposition \ref{liminfbound}).
We then deduce that 
\begin{equation}\label{accure1}
\#\{ T\in  {\cal S}^s_\e: T\not\subset {\bf T}_\e\}\le c{\rho\over\e}\delta   
\end{equation}
and, in addition, there exist $c{\rho\over\e}$ disjoint polygonal chains joining $P^\e_-$ and the side
of $Q^{\nu_0}_\rho(x_0)$ lying on $\{y:\langle y-x_0,\nu\rangle=-\rho/2\}$ such that they do not intersect ${\cal S}^s_\e$. 
Since $v_{\e,\rho}\to\widetilde u$, arguing as in \eqref{prit}, we deduce that 
\begin{equation}\label{dubbio}
\int_{P^\e_\pm} |u_\e- u^\pm(x_0)|d{\cal H}^1= \rho\, o_\e(1)+o(\rho)
\end{equation}
 as $\e\to 0$.

For fixed $\rho>0$ we may now consider the scaled functions
$$
w^\rho_\e(y)= {1\over\rho}u_\e(x_0+\rho y)=
{1\over\rho}v_{\e,\rho}(y).
$$
We may still argue as in Proposition \ref{compactness} and deduce that $w^\rho_\e$ converge to
$u^\rho(y)={1\over\rho}u(x_0+\rho y)$. Note that $\nabla u^\rho(y)=\nabla u(x_0+\rho y)$.

For fixed $\rho$, we may now consider the functions
$$
\tilde w^\rho_\e =w^\rho_\e\chi_{Q^{\nu_0}\setminus{1\over\rho}{\cal S}^s_\e},
$$
which are now discontinuous at $\partial {\cal S}^s_\e$.
They satisfy

(i)  the determinant of $\nabla w^\rho_\e$ is equibounded, by the definition of ${\cal S}^s_\e$;

(ii) $\nabla w^\rho_\e$ are equibounded;

(iii) ${\cal H}^1( \partial {\cal S}^s_\e)$ is bounded;

(iv) the measures ${\partial w^\rho_\e\over\partial\nu^\perp}{\cal H}^1\LL \partial {\cal S}^s_\e$
are equibounded, by (ii) and since  $|{\partial w^\rho_\e\over\partial\nu^\perp}|\le 2s$ by definition.
Here and below we simply denote by $\nu$ the normal to a discontinuity set without specifying 
of which function, which is clear from the context.


By Theorem 5.8 in \cite{ABG} from (i)--(iv) above and the convergence 
$\tilde w^\rho_\e\to u^\rho$ we deduce that
the measures ${\partial w^\rho_\e\over\partial\nu^\perp}{\cal H}^1\LL \partial {\cal S}^s_\e$ converge to ${\partial u^{\rho,\delta}\over\partial\nu^\perp}{\cal H}^1\LL S_{u^{\rho}}$. We choose the orientation of $\nu^\perp$ so that the determinant constraint on $u_\e$ can be rewritten on each segment of $P^\e_\pm$ as
$$
\Bigl\langle{\partial u_\e\over\partial\nu^\perp}, (u_\e^+-u_\e^-)^\perp\Bigr\rangle>0,
$$
where $u_\e^-$ and $u_\e^+$ are the values on two vertices of the corresponding triangle, with $u_\e^-\in P^\e_-$ and $u_\e^+\in P^\e_+$. We then have, by \eqref{dubbio}
\begin{eqnarray*}
0&\le& \int_{Q^{\nu_0}_\rho(x_0)\cap P_-^\e} \Bigl\langle {\partial u_\e\over\partial\nu^\perp}, (u_\e^+-u_\e^-)^\perp\Bigr\rangle d{\cal H}^1
\\
&=& \int_{Q^{\nu_0}_\rho(x_0)\cap P_-^\e} \Bigl\langle {\partial u_\e\over\partial\nu^\perp}, (u^+(x_0)-u^-(x_0))^\perp\Bigr\rangle d{\cal H}^1 +\rho\, o_\e(1)+ o(\rho)
\\
&=& \rho \int_{Q^{\nu_0}\cap {1\over \rho}(P_-^\e-x_0)} \Bigl\langle {\partial w^\rho_\e\over\partial\nu^\perp}, (u^+(x_0)-u^-(x_0))^\perp\Bigr\rangle d{\cal H}^1 +\rho\, o_\e(1)+ o(\rho)
\\
&=& \rho \int_{Q^{\nu_0}\cap S_{u^\rho}} \Bigl\langle {\partial u^\rho\over\partial\nu^\perp}, (u^+(x_0)-u^-(x_0))^\perp\Bigr\rangle d{\cal H}^1+
\rho\, o_\e(1)+ o(\rho)
\\
&=& \rho\,\langle R^-\nu^\perp, (u^+(x_0)-u^-(x_0))^\perp\rangle 
+\rho\, o_\e(1)+ o(\rho)\\
&=& \rho\,\langle R^-\nu, u^+(x_0)-u^-(x_0)\rangle 
+\rho\, o_\e(1)+ o(\rho).
\end{eqnarray*}
Dividing by $\rho$ and letting $\e,\rho\to0$ we obtain the first claim in \eqref{bocia}.
The analogous inequality with $R^+\nu$ is obtained in the same way.\endproof

\begin{remark}[conjecture]\rm We have obtained a weaker necessary condition than the sufficient ones 
in \eqref{incompen2}. We conjecture that this is only a technical issue and indeed those condition are also necessary in order that the energy density be minimal on $S_u$.
\end{remark}

\section{Small deformations}
In the previous sections we have showed how the ``positive-determinant'' constraint on the fracture is not closed, and must be substituted by complex non-local energetic considerations. The constraint simplifies in the case of small deformations; i.e., if we require that 
\begin{equation}
u= \hbox{id} +\delta_\e v
\end{equation}
(or any function with gradient in $SO(2)$ in place of the identity) with $\delta<\!<1$. 

If $\sqrt\e<\!<\delta_\e$ then we can still apply the compactness argument as above to obtain, from the boundedness of $E_\e(u_\e)$, that $v$ determines a partition into sets of finite perimeter. Since $u^+(x)- u^-(x)= \delta_\e(v^+(x)- v^-(x))$
and $R^\pm=\hbox{id}+o(1)$, in the case of a single coordinate normal $\nu$ the constraint on $u$
$$
\langle u^+(x)- u^-(x), R^\pm \nu \rangle\ge 0\,
$$
can be translated into one on $v$ that reads 
\begin{equation}
\langle v^+(x)- v^-(x), \nu \rangle\ge 0
\end{equation}
(in the case of a general non-coordinate normal we have to take into account the microscopic oscillations of $S_v$ as usual). Note that this is the usual {\em infinitesimal opening fracture constraint}, which is a closed constraint.

\begin{remark}[linearized regime]\rm
The case $\delta_\e$ of the order of $\sqrt \e$ corresponds to the one studied by Braides, Lew and Ortiz in the one-dimensional case when bulk and surface terms have the same scaling. Hence, the compactness argument used above does not hold, and the limit energy possesses both a bulk and a surface term. The linearization of the bulk term has been studied by Braides, Solci and Vitali \cite{BSV}, and can be used together with the analysis above.
The resulting $\Gamma$-limit can be explicitly written as
$$
\int_\Omega {\bf C}\nabla v:\nabla v\dx+ \int_{S_v}\varphi(\nu_v)d\Huno
$$
on piecewise $H^1$ functions whose discontinuity set is a union of smooth lines, $v$ satisfies the infinitesimal opening fracture constraint on $S_v$. Note that a construction as in Section \ref{triplecon} implies that the condition on triple points can be dropped altogether (since microfractures can be chosen vanishing with $\delta_\e$). The proof of the upper bound for the $\Gamma$-limit is a technically complex matter due to the lack of density theorems for $v$ satisfying the resulting constraints, and will not be addressed here. Related results can be found in works by Friedrich and Schmidt
\cite{FS1,FS2,FS3,F1}. 
\end{remark}

\section{Conclusions}
We have examined a two-dimensional system of nearest-neighbour interactions for Lennard Jones pair interactions parameterized on a triangular lattice, with a microscopic positive-determinant constraint which mimics the effect of long-range interactions and limits the ground states to rotations. The goal is to exhibit a continuum approximation so as to test the validity of such a simplified model.

In parallel to the one-dimensional case we have focused our attention on the fracture term of the resulting continuum approximation by suitably scaling the energies. We have thus determined

$\bullet$ a surface energy density $\varphi$ which reflects the triangular symmetries of the underlying lattice;

$\bullet$ conditions on the interface that reflect the positive-determinant constraint. The conditions on the interfaces are of a novel type that take into account both the gradient of the deformation on both sides of the fracture and the orientation of the fracture site in the reference configuration. In addition we also have a positive-determinant constraint on points where more fracture meet (triple points);

$\bullet$ such conditions are not a closed constraint, and can be removed by adding ``fictitious'' micro-fractures. The optimal location and form of those micro-fractures depends on the corresponding macroscopic deformation, and is a complex optimization problem;

$\bullet$ in the case of small deformations the conditions on the interfaces reduce to the infinitesimal opening fracture constraint, which actually suggests a linearized approximation in the spirit of Braides, Lew and Ortiz with that constraint on the fracture (also analyzed in recent works by Friedrich and Schmidt \cite{FS1,FS2,FS3,F1}).

We have neglected interactions other than nearest neighbours. In this way the interfacial energy does not reflect the possibility of surface relaxation; i.e., the fact that atomistic interactions are unbalanced close to interfaces. This is an important effect, especially at the boundary of the domain and in the determination of the location of fractures. It has been partially addressed by Theil \cite{Th2}, and should be included in further investigations on the subject. The one-dimensional analysis by Braides and Solci \cite{BSo} suggests that considering longer range of interactions (e.g., next-to-nearest neighbours) could be used
in the place of the determinant constraint in order to eliminate non-local effects for non-opening cracks,
at the expense of introducing internal and external boundary layers.

\section*{Appendix: positive vs non-negative determinant constraint}
We have considered a {\em strictly positive} microscopic determinant constraint. Strictly positive inequalities are weakened in the limit; however, our choice of not directly considering
weak inequalities allows to rule out some  additional ``unphysical'' deformations which would have to be taken into account by directly considering a non-negative microscopic determinant constraint. This would correspond to allowing microscopic interpolations to ``collapse'' triangles to segments on the jump set even though such collapsed triangles cannot be viewed as a limit for the strictly positive determinant case.
This may happen in the case of rotations $R^\pm\in SO(2)$ on both sides of the interface with $R^+=-R^-$. 
\begin{figure}[htbp]
\begin{center}
\includegraphics[width=.33\textwidth]{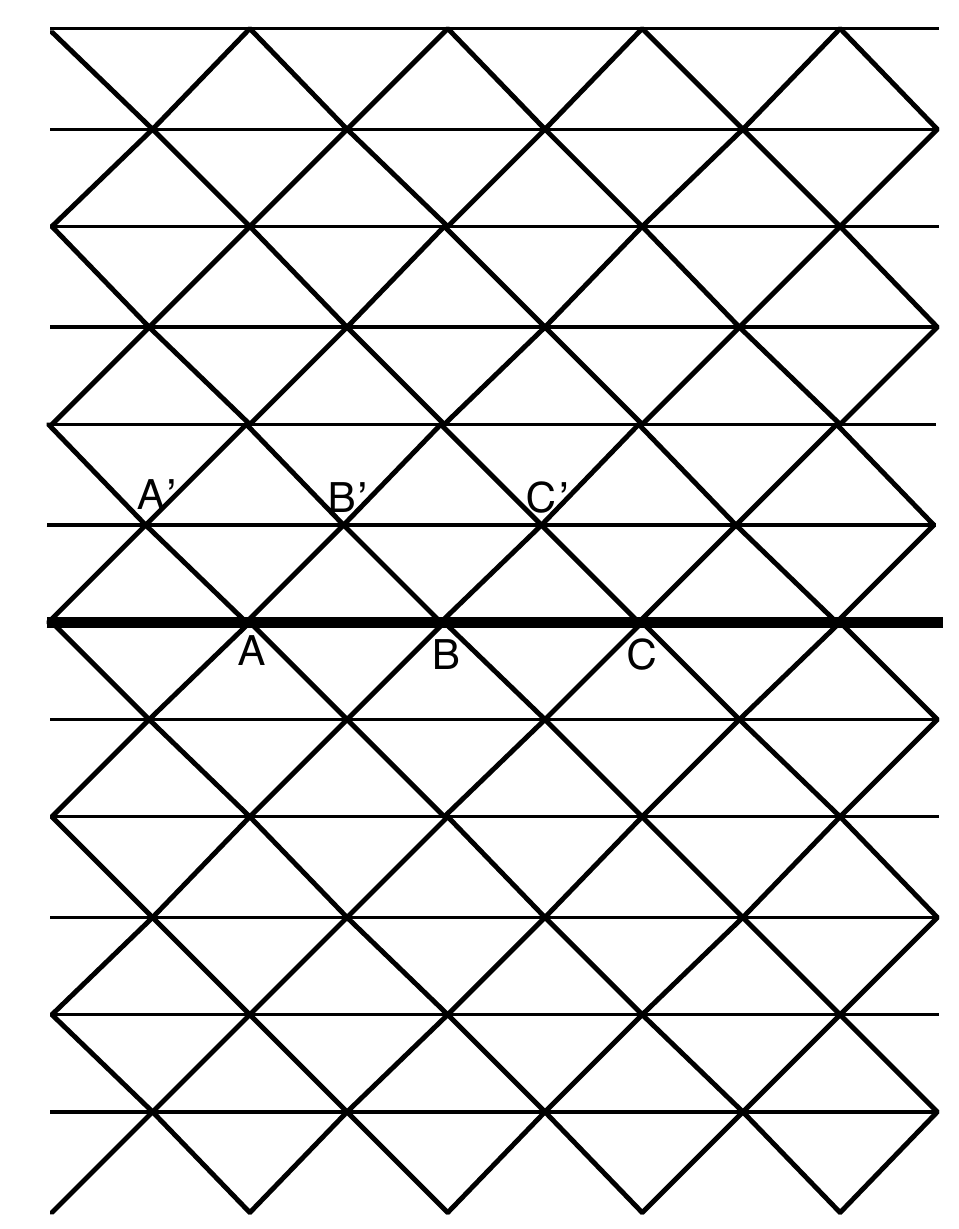}\qquad
\includegraphics[width=.44\textwidth]{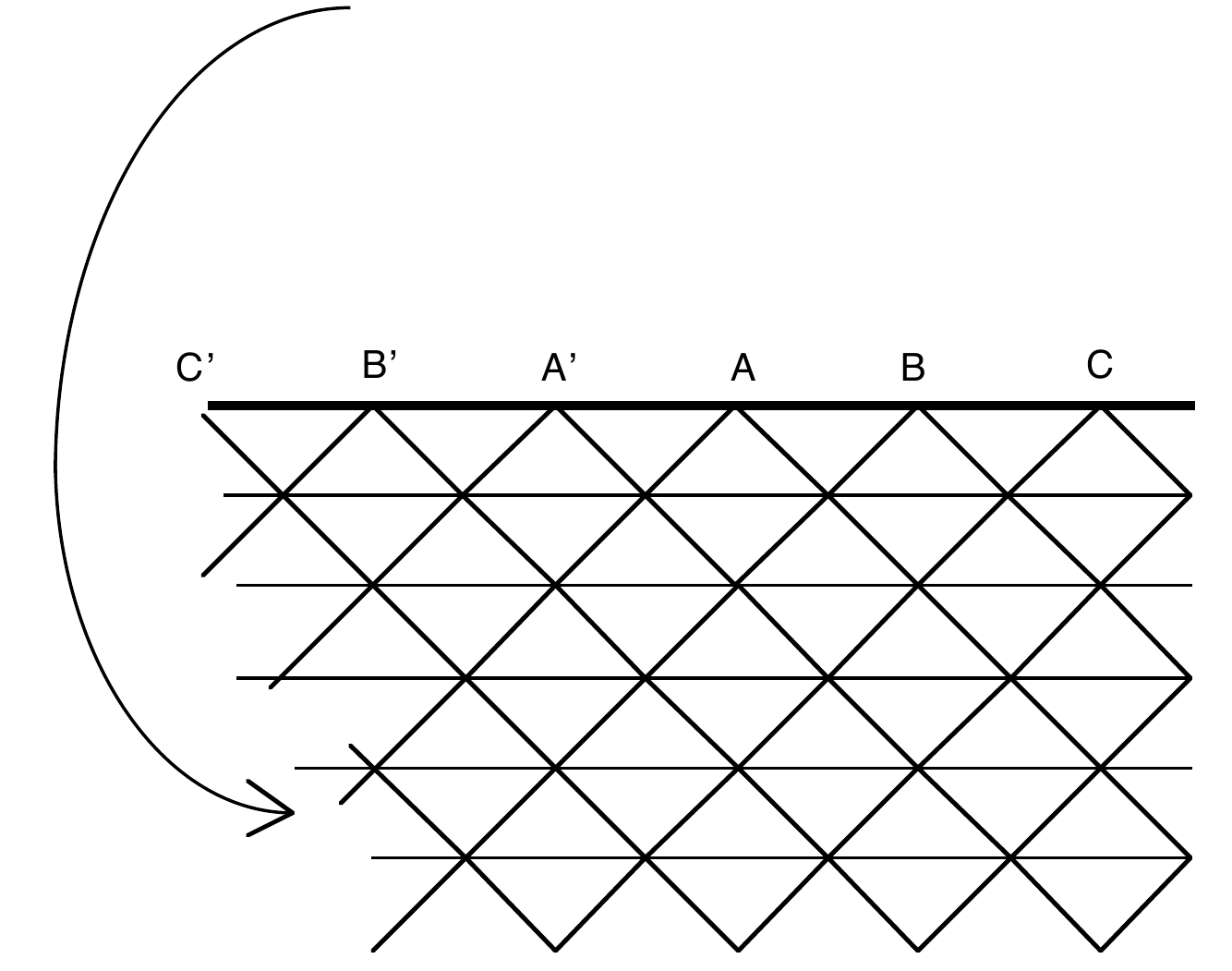}
\caption{Zero-determinant fracture with a planar interface and a $180$-degree rotation -- reference and deformed microscopic configurations}
\label{zerodet1-2}
\end{center}
\end{figure}
In Fig.~\ref{zerodet1-2} we depict such a microscopic deformation along a single coordinate line, where all points of two rows are aligned. Note that allowing zero-determinant deformations would include such a macroscopic deformation in the set of ``minimal interfacial energy'', while in our setting the same must be achieved by the introduction of at least one extra layer of atoms, thus doubling the energy.

Another possibility for having $R^+=-R^-$ is with the partition composed of a pair of supplementary angles as in Figure \ref{zerodet3-4}. In this case it is possible to ``rotate'' one of the two angles by $\pi$ with positive or zero determinant on the deformed triangles along the discontinuity set.
In both cases, however, we have a macroscopical failure of impenetrability, so we have regarded these cases as degenerate by considering only the strictly-positive-determinant constraint.

\begin{figure}[htbp]
\begin{center}
\includegraphics[width=.33\textwidth]{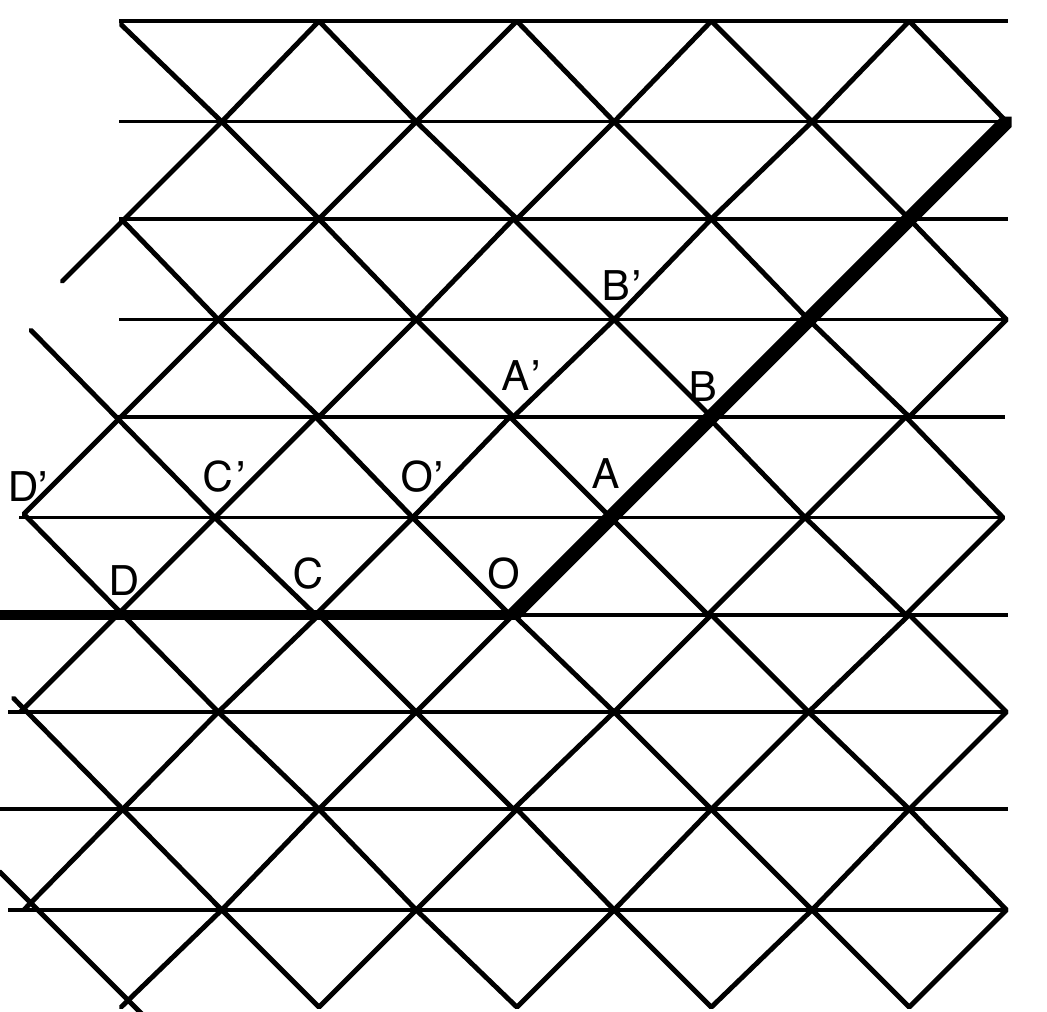}\qquad
\includegraphics[width=.44\textwidth]{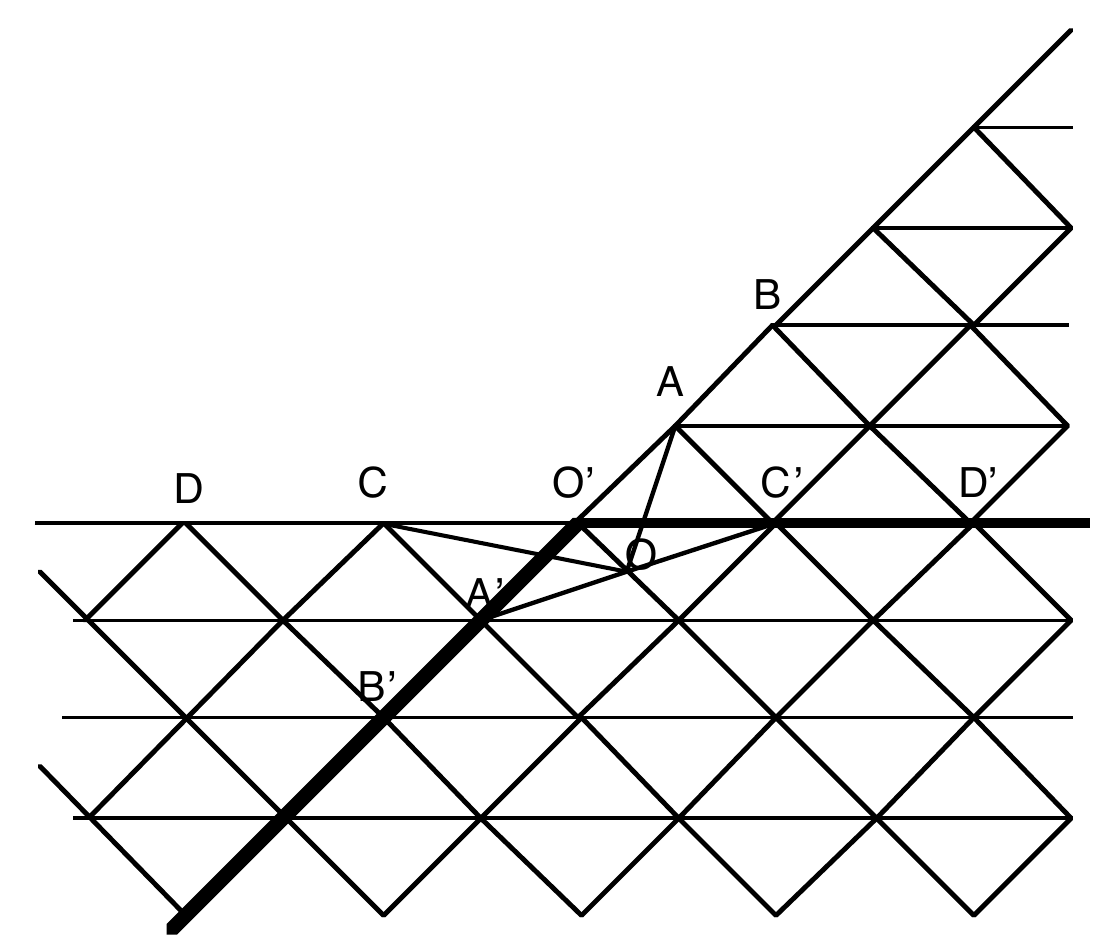}
\caption{Zero-determinant fracture with an angular interface and a $180$-degree rotation -- reference and deformed microscopic configurations}
\label{zerodet3-4}
\end{center}
\end{figure}


\end{document}